%% file: main.tex
\documentclass[10pt,twocolumn,letterpaper]{article}
\input{misc/math_commands.tex}

\usepackage{xcolor}

\usepackage[pagenumbers]{cvpr} 
\input{misc/preamble}
\definecolor{cvprblue}{rgb}{0.21,0.49,0.74}
\usepackage[pagebackref, breaklinks, colorlinks,
    linkcolor=cvprblue,
    citecolor=cvprblue,
    urlcolor=cvprblue
]{hyperref}

\PassOptionsToPackage{hyphens}{url}
\usepackage{diagbox, graphicx}
\usepackage{algorithm, algpseudocode, setspace}
\usepackage{amssymb, amsmath, amsthm}
\usepackage{multirow, pifont, caption}
\usepackage{lipsum}

\newcommand{\cmark}{\textcolor{green!60!black}{\ding{51}}}
\newcommand{\xmark}{\textcolor{red!70!black}{\ding{55}}}

\usepackage[table]{xcolor}
\usepackage{wrapfig, sidecap, adjustbox, float}
\usepackage{booktabs, thmtools}

\usepackage{arydshln}

\def\x{{\mathbf x}}
\def\z{{\mathbf z}}
\def\u{{\mathbf u}}
\def\y{{\mathbf y}}

\def\n{{\mathbf n}}

\title{PnP-CM: Consistency Models as Plug-and-Play Priors for Inverse Problems}

\author{Merve G\"{u}lle$^*$ \qquad Junno Yun$^*$ \qquad Ya\c{s}ar Utku Al\c{c}alar$^*$ \qquad Mehmet Ak\c{c}akaya \vspace{1mm}\\
University of Minnesota\\{\tt\small \{glle0001, yun00049, alcal029, akcakaya\}@umn.edu}\\$^*$\small Equal contribution\\\url{https://github.com/MerveGulle/PnP-CM}}

\begin{document}

\twocolumn[{
\renewcommand\twocolumn[1][]{#1}
\maketitle
\vspace{-2em}
\begin{center}
    \centering
    \captionsetup{type=figure}
    \includegraphics[width=0.9\linewidth]{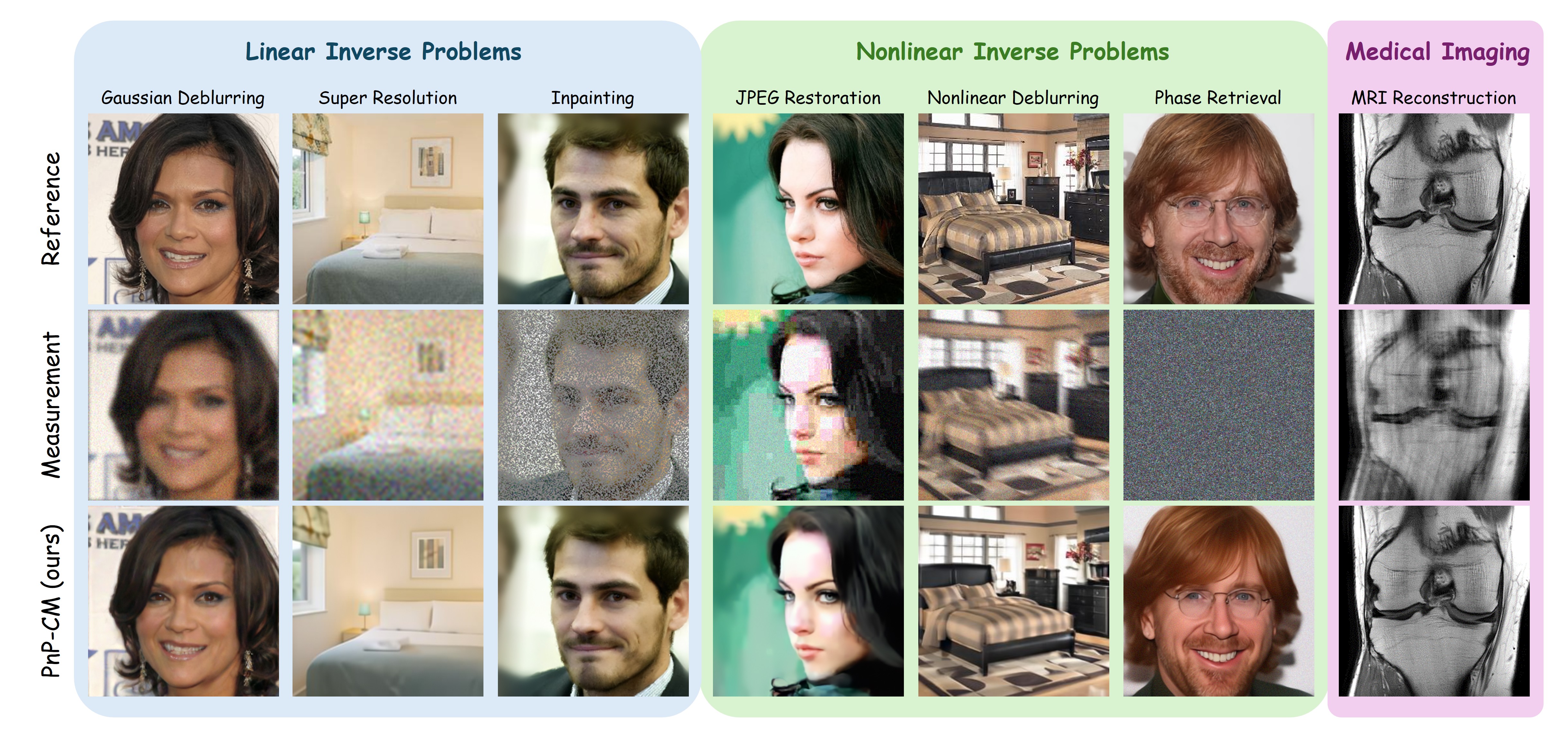}
    \vspace{-2mm}
    \caption{Representative results of our method (PnP-CM) across a diverse set of inverse problems. \textbf{Left to right:} \emph{linear inverse problems} (Gaussian deblurring, super-resolution, inpainting) and \emph{nonlinear inverse problems} (JPEG artifact removal, nonlinear deblurring, phase retrieval), all with additive Gaussian noise $\sigma=0.05$, followed by \emph{MRI reconstruction} with inherently noisy measurements. \textbf{Top to bottom:} reference, degraded measurement, and PnP-CM reconstruction.}
    \label{fig:overview}
    \vspace{1mm}
\end{center}}]

\looseness=-1
\begin{abstract}
Diffusion models have found extensive use in solving inverse problems, by sampling from an approximate posterior distribution of data given the measurements. Recently, consistency models (CMs) have been proposed to directly predict the final output from any point on the diffusion ODE trajectory, enabling high-quality sampling in just a few neural function evaluations (NFEs). CMs have also been utilized for inverse problems, but existing CM-based solvers either require additional task-specific training or utilize data fidelity operations with slow convergence, limiting their applicability to large-scale problems and making them difficult to extend to nonlinear settings. In this work, we reinterpret CMs as proximal operators of a prior, enabling their integration into plug-and-play (PnP) frameworks. Specifically, we propose {\bf PnP-CM}, an ADMM-based PnP solver that provides a unified framework for solving a wide range of inverse problems, and incorporates noise perturbations and momentum-based updates to improve performance in the low-NFE regime. We evaluate our approach on a diverse set of linear and nonlinear inverse problems. We also train and apply CMs to MRI data for the first time. Our results show that PnP-CM achieves high-quality reconstructions in as few as 4 NFEs, and produces meaningful results in 2 steps, highlighting its effectiveness in real-world inverse problems while outperforming existing  CM-based approaches.
\end{abstract}

\section{Introduction}
Diffusion models (DMs) have established themselves as state-of-the-art generative models, capable of synthesizing high-fidelity and diverse samples across a wide range of domains~\citep{ho2020ddpm,song2021ddim,song2021scoreSDE,dhariwal2021beatGANs,karras2022elucidating,yun2025_no_REPA}. Beyond unconditional generation, DMs have also been widely adopted as powerful priors for solving imaging inverse problems~\citep{chung2023dps,song2023pgdm,wang2023ddnm,alcalar2025_CAMSAP_ZADS}, where the goal is to recover clean images from degraded measurements. Conventional DM-based solvers typically combine the unconditional score function with a data fidelity term tied to the forward operator, and generate reconstructions by integrating either the reverse stochastic differential equation (SDE) or the probability flow ordinary differential equation (ODE)~\citep{song2021scoreSDE}. While highly effective, these approaches are computationally expensive, often requiring hundreds of neural function evaluations (NFEs) to achieve competitive results, which limits their practicality in large-scale or time-sensitive applications~\citep{salimans2022progressive, song2023_consistency_models}.

Several approaches have been proposed to accelerate DM sampling, including progressive distillation~\citep{salimans2022progressive,meng2023distillation}, rectified flow~\citep{liu2023flow,liu2024instaflow}, and distribution matching distillation (DMDs)~\citep{yin2024DMDs,yin2024improved}. Among these, consistency models (CMs)~\citep{song2023_consistency_models, lu2025simplifying} distill the generative power of DMs into a consistency function that maps any point on the diffusion ODE trajectory directly to its clean origin. This design enables high-quality sampling in as few as one to four NFEs, which has made CMs particularly attractive as efficient priors for inverse problems. However, existing CM-based methods face limitations. For instance, CoSIGN~\citep{zhao2024cosign} incorporates an external ControlNet~\citep{zhang2023ControlNet} to encode measurement operators, requiring task-specific training for each degradation. This restricts its generalization across inverse problems, since adapting to a new measurement operator typically entails retraining or fine-tuning the ControlNet. Similarly, CM4IR~\citep{garber2025CM4IR} leverages a form of pseudo-inverse guidance for measurement consistency, a strategy fundamentally tied to the conditioning of the forward operator. Thus, its use in highly ill-conditioned systems is nontrivial, as the pseudo-inverse becomes unstable, thereby limiting its reliability in more complex imaging tasks.

\looseness=-1
In this work, we propose \textbf{PnP-CM}, a novel framework that reinterprets CMs as proximal operators within a learned prior setting. This perspective allows CMs to be seamlessly embedded into plug-and-play (PnP) optimization~\citep{venkatakrishnan2013plug}, combining their sampling efficiency with the convergence guarantees of PnP methods. Our solver builds on the alternating direction method of multipliers (ADMM), with enhancements to speed-up convergence while preserving theoretical guarantees \cite{chan2016_PnP_ADMM}. Representative results of our method are shown in \figref{fig:overview}. Our main contributions are: 
\begin{itemize}[leftmargin=*, itemsep=0em, topsep=0pt]
    \item We introduce \textbf{PnP-CM}, a PnP framework that reinterprets CMs as proximal operators of a prior. We further enhance the runtime of PnP-CM using controlled noise injection and momentum, while showing that they maintain the convergence properties of the baseline algorithm.
    \item We evaluate PnP-CM on natural image datasets such as LSUN Bedroom~\citep{yu2015lsun} and CelebA-HQ~\citep{karras2018progressive} across a range of linear and nonlinear inverse problems, achieving state-of-the-art results in as few as four NFEs and producing meaningful outputs in just two steps, consistently outperforming DM- and CM-based comparison baselines in both quality and efficiency.
    \item We demonstrate that CMs can be trained and applied to large-scale medical imaging for practical MRI reconstruction. By training a CM on fastMRI database~\citep{knoll2020fastmri_dataset-journal}, \emph{for the first time}, we show that PnP-CM achieves high-quality reconstructions in 4 NFEs, outperforming existing diffusion-based methods in both quality and speed.
\end{itemize}

\section{Background and Related Work}
\subsection{Diffusion Models}
DMs~\citep{ho2020ddpm,song2021scoreSDE,karras2022elucidating} define a generative process by progressively corrupting data with Gaussian noise and then learning how to invert this process. Let $p_{\text{data}}(\x)$ denote the data distribution. The forward noising process is often described by the stochastic differential equation (SDE)~\citep{song2021scoreSDE}
\begin{equation}
    \mathrm{d}\x_t = \mu(\x_t, t)\,\mathrm{d}t + \sigma(t)\,\mathrm{d}w_t, \quad t \in [0,\mathrm{T}],
    \label{eq:sde}
\end{equation}
where $w_t$ is standard Brownian motion, and $\mu$ and $\sigma$ specify the drift and diffusion schedules. This ensures that the marginal distribution of $\x_t$, denoted $p_t(\x)$, evolves smoothly from $p_0(\x)=p_{\text{data}}(\x)$ to a tractable prior $\pi(\x)$ as $t \to \mathrm{T}$ for large $\mathrm{T}$, where $\pi(x)$ is typically chosen as an isotropic Gaussian distribution. An important observation is that the SDE in \eqnref{eq:sde} is associated with a deterministic probability flow ODE~\citep{song2021scoreSDE},
\begin{equation}
    \mathrm{d}\x_t = \Big[\,\mu(\x_t, t) - \tfrac{1}{2}\sigma^2(t)\,\nabla \log p_t(\x_t)\Big] \mathrm{d}t, \label{eq:pf_ode}
\end{equation}
which shares the same marginal distributions as the SDE. Because the drift term involves the score function $\nabla \log p_t(\x)$, diffusion models are often viewed as score-based generative models. In practice, one learns a neural score network $s_\phi(\x,t)$ to approximate $\nabla \log p_t(\x)$ using denoising score matching. To generate new samples, this estimate is substituted into \eqnref{eq:pf_ode}. Then  $\x_\mathrm{T} \sim \pi(\x)$ is sampled, and \eqnref{eq:pf_ode} is integrated backwards in time to $t$ = 0 using a numerical solver such as Euler~\citep{song2021ddim,song2021scoreSDE} or Heun’s~\citep{karras2022elucidating} method. The terminal point $\x_0$ is then regarded as a sample from $p_{\text{data}}(\x)$. For stability, the integration is typically stopped at a small $\varepsilon > 0$ instead of exactly at zero.

\subsection{Consistency Models}
CMs~\citep{song2023_consistency_models,lu2025simplifying} were proposed to accelerate diffusion sampling while preserving sample quality. Instead of sequentially denoising through the entire diffusion trajectory, CMs learn a mapping that directly projects a noisy sample $\x_t$ at any timestep $t$ to its corresponding clean origin $\x_0$. Formally, a consistency function $f_\theta(\x_t, t)$ is trained to satisfy
\begin{equation}
    f_\theta(\x_t, t) = f_\theta({\bf x}_{t'}, t'), \quad \forall t,t' \in [\epsilon,\mathrm{T}],
\end{equation}
with the boundary condition $f_\theta(\x_{\epsilon}, \epsilon) = \x_\epsilon$. Here $\x_t$ and $\x_{t'}$ are drawn from the forward process defined in \eqnref{eq:sde}. Training can be carried out either via consistency distillation (CD) by distilling the behavior of an existing DM or by consistency training (CT) which learns $f_\theta$ directly from data. In both cases, the objective enforces that predictions from different noisy inputs of the same target agree, leading to the consistency loss
\begin{equation}
    \mathcal{L}_{\text{CM}} = \mathbb{E}_{\x_0, t, \x_t, t', \x_{t'}}\big[w(t)d(f_\theta(\x_t, t), f_{\theta^{-}}(\x_{t'}, t'))\big],
\end{equation}
where $(\x_t, t)$ and $(\x_{t'}, t')$ are two independently sampled noised versions of $\x_0$ at timesteps $t$ and $t'$, $w(t)$ is a weighting function, $d(\cdot, \cdot)$ is a distance function such as $\ell_2$ or LPIPS, and $f_{\theta^-}$ describes a corresponding ``teacher'' network achieved by exponential moving average (EMA). At inference, sampling does not need to begin from pure Gaussian noise. Instead, CMs allow initialization from any intermediate state $\x_t$, and the consistency function maps it directly toward the clean estimate:
\begin{equation}
    \hat{\x}_0 = f_\theta(\x_t, t), \quad t \in (0,\mathrm{T}].
\end{equation}
In practice, one may start from $\x_\mathrm{T} \sim \pi(\x)$, where $\pi(\x)$ denotes the prior distribution at time $\mathrm{T}$ or from a measurement-dependent initialization, and apply $f_\theta$ either once or over a few discretized steps for improved stability. This flexibility enables high-quality image generation in as few as 1–4 NFEs, in stark contrast to the hundreds typically required by diffusion-based methods.

\subsection{Generative Approaches to Inverse Problems}
Inverse problems aim to recover an unknown signal $\x \in \mathbb{R}^n$ from a set of degraded observations $\y \in \mathbb{R}^m$. The measurement process is typically modeled as
\begin{equation} \label{eq:inverse}
    \y = \mathcal{A}(\x) + \n, 
\end{equation}
where $\mathcal{A}(\cdot)$ is the (possibly nonlinear) forward operator, which reduces to a matrix $\mathbf{A} \in \mathbb{R}^{m \times n}$ in the linear case, and $\n$ is measurement noise.
For $m < n$, the problem is ill-posed, and a common strategy to use maximum a posteriori (MAP) estimation:
\begin{equation} \label{eq:MAP}
    \hat{\x} 
    = \displaystyle \arg\max_{\x} p(\x | \mathbf{y}) \\
    = \displaystyle \arg\min_{\x} - \log p(\mathbf{y} | \x) - \log p(\x),
\end{equation}
where the likelihood term $-\log p(\mathbf{y}| \x)$ is the \emph{data fidelity} and the prior term $-\log p(\x)$ captures information about the image statistics.

\paragraph{DM-based approaches} These adapt DMs to approximate the posterior distribution $p(\x|\y)$ by combining score-based priors with data fidelity terms~\citep{chung2023dps,graikos2022diffusion,zhu2023DiffPIR,xu2024_DPnP,wu2024_PnP-DM,park2025plug,song2023pgdm,mardani2024reddiff,alcalar2024ZAPS,moufad2025midpointguidance}. Among these, diffusion posterior sampling (DPS)~\citep{chung2023dps} modifies the denoising step by incorporating the gradient with respect to $\x_t$ of the measurement likelihood evaluated at the denoised Tweedie estimate of the clean data, as an approximation to $\log p({\bf y}|\x_t)$. Intuitively, this adjustment encourages each denoising update to remain consistent with the observed measurements while following the learned data prior. While effective, DPS typically requires hundreds of NFEs to achieve stable reconstructions, which severely limits its practicality. Another line of work integrates the classical PnP framework directly into the diffusion sampling process by decoupling the prior and data terms~\cite{graikos2022diffusion,zhu2023DiffPIR,xu2024_DPnP,wu2024_PnP-DM,park2025plug}. The DM acts as a generative denoiser for the prior subproblem, while the data fidelity is enforced analytically via a proximal operator. While these approaches highlight the promise of combining generative models with PnP formulations, they still typically require on the order of 100 NFEs, limiting their practical efficiency. More recent approaches~\citep{song2023pgdm,mardani2024reddiff,alcalar2024ZAPS,moufad2025midpointguidance} have further reduced the computational cost to roughly 20–100 NFEs. However, this is still far from the efficiency required for large-scale or real-time applications.

\paragraph{CM-based solvers} CMs~\citep{song2023_consistency_models,lu2025simplifying} offer another direction by distilling diffusion trajectories into a direct mapping from noisy states to clean reconstructions, enabling sampling in as few as 1–4 NFEs. CoSIGN~\citep{zhao2024cosign} further integrates a ControlNet to encode forward operators, but requires retraining or fine-tuning for each new measurement model. CM4IR~\citep{garber2025CM4IR} instead combines CMs with back-projection using a pseudo-inverse to enforce measurement consistency, a formulation that is effective for some linear operators, but does not readily extend to highly ill-conditioned cases without additional heuristics or to non-linear inverse problems. These limitations underscore the need for CM-based methods that are general and efficient.

\section{PnP-CM: A Consistency Model-Based Plug-and-Play Framework for Inverse Problems}
In this work, we propose PnP-CM, a plug-and-play framework that interprets CMs as proximal operators of an image prior, enabling efficient inverse problem solving with only a few NFEs. We instantiate the framework within an ADMM scheme and incorporate established acceleration strategies, such as noise injection~\cite{song2023_consistency_models} and momentum~\citep{nesterov1983method,goldstein2014fast,deng2025_SM_ADMM}, while preserving convergence guarantees. 

\subsection{Using CMs in the PnP Framework}
An alternative optimization-centric way to view the MAP estimation of \eqnref{eq:MAP} is:
\begin{equation}\label{eq:OPT}
    \arg\min_{\x} f(\x) + \lambda g(\x),
\end{equation}
where $f(\cdot)$ is a data fidelity term, while $g(\cdot)$ is a regularizer/prior~\citep{tibshirani1996regression, wipf2004sparse, bishop2006pattern, park2008bayesian}. In particular, for additive \iid Gaussian noise, commonly encountered in computational imaging problems~\citep{chung2023dps, hammernik2023SPM, akcakaya2022_SPMsurvey, junno2025_TE-MRI_NIPS,alcalar2025_CUPID}, $f(x) = \frac{1}{2} ||{\bf y}-{\bf Ax}||_2^2.$ ADMM solves \eqnref{eq:OPT} by variable splitting with an augmented Lagrangian penalty, and alternating minimization:
\begin{align}
    \z^{(k+1)} &= \arg\min_{\z} \; f(\z) + \tfrac{\rho}{2}\|\z-\x^{(k)}+\u^{(k)}\|_2^2, \label{eq:z_update} \\
    \x^{(k+1)} &= \arg\min_{\x} \; g(\x) + \tfrac{\rho}{2}\|\z^{(k+1)}-\x+\u^{(k)}\|_2^2, \label{eq:x_update} \\
    \u^{(k+1)} &= \u^{(k)} + \z^{(k+1)}-\x^{(k+1)},\label{eq:u_update} 
\end{align}
where the $\z$-update enforces data fidelity, the $\x$-update applies the proximal operator associated with the prior, and $\u$ is the dual variable~\citep{boyd2011admm, hong2016convergence, eckstein2015understanding}. In practice, the proximal operator of $g(\cdot)$ is often intractable. PnP methods~\citep{venkatakrishnan2013plug, chan2016_PnP_ADMM, kamilov2017plug} overcome this by replacing the proximal update with an off-the-shelf denoiser $D_\sigma(\cdot)$:
\begin{equation} \label{eq:PnP-den}
    \x^{(k+1)} = D_{\sigma_k}\!\left({\bf z}^{(k+1)}+\u^{(k)}\right).
\end{equation}
Such denoisers can also be based on deep neural networks, leveraging their ability to learn rich image priors for improved restoration quality~\citep{chan2016_PnP_ADMM, chan2019performance}.

\looseness=-1
One of the other advantages of the ADMM baseline is related to the quadratic penalty in \eqnref{eq:z_update}. This effectively results in Tikhonov regularization for the data fidelity, improving the conditioning of the minimization sub-problem, making it less sensitive to the conditioning of the forward operator compared to methods based on proximal gradient descent (PGD)~\citep{DengYin2016, hong2017convergence}. In fact, existing CM-based methods, such as CM4IR~\citep{garber2025CM4IR} can be seen as a type of preconditioned PGD with data fidelity based on a back-projection objective, $\frac12 ||{\bf A}^\dagger ({\bf y - Ax})||_2^2$~\citep{TirerGiryes2020}, which inherits the dependencies of the baseline algorithm on the conditioning of the forward operator. Furthermore, it is well-understood that ADMM requires fewer iterations compared to PGD in practice for convergence to modest accuracy, sufficient for imaging inverse problems~\citep{boyd2011admm}.

In this study, we use CMs as denoisers within the PnP framework in \eqnref{eq:PnP-den}. While this formulation provides an effective approach to inverse problems, directly using CMs does not fully realize their potential in the low-NFE regime. To better align the iterative updates with the noise-level dependence of CMs and improve the performance within a limited number of outer iterations, we incorporate previously-studied techniques, such as controlled noise perturbations and momentum-based updates, as described next.

\input{Algos/pnp_cm_algo}

\subsection{Acceleration in the Low-NFE Regime}
Since our goal is to perform CM-based reconstruction in as few as 2 NFEs, we focus on improving performance in the low-iteration regime of PnP-CM. In particular, we consider modifications that enable high-quality solutions within a small number of iterations, while preserving the convergence behavior of the underlying PnP formulation.

\paragraph{Noise injection} A common strategy in CMs is noise injection to improve generative quality~\citep{song2023_consistency_models, song2024improved, lu2025simplifying, garber2025CM4IR}. This idea has also been explored in the context of image restoration and inverse problems for a longer time~\citep{BM3Dicip2007,atchade2017perturbed}. In broader optimization community, noise injection was shown to help iterative optimization methods escape saddle points and explore the solution space more effectively~\citep{huang2021escaping, guo2020escaping}. We build on the ideas from these works, and add controlled perturbations for the CM to operate at higher noise levels, improving its performance in the limited iteration regime. Importantly, this modification is compatible with standard PnP formulations. To make this precise, we analyze the effect of noise injection within the PnP-ADMM framework~\cite{chan2016_PnP_ADMM}, which serves as a convenient setting for studying convergence:

\begin{restatable}{theorem}{noiseinjection}
\label{cor:noise_injection}
Consider the PnP-ADMM algorithm, where the proximal operator is replaced by an $L$-Lipschitz continuous denoiser $D_{\sigma}$. Suppose noise injection is applied to the denoiser input, \ie
\begin{equation}
\x^{(k+1)} = D_{\sigma}(\z^{(k+1)}+\u^{(k)}+\boldsymbol{\eta}_k), \nonumber
\end{equation}
where $\x$ and $\z$ are the primal variables, $\u$ is the dual variable, $\boldsymbol{\eta}$ is an injected noise term and $k$ counts the iterations. If the PnP-ADMM algorithm without noise injection converges to a fixed point, then the algorithm with noise injection still converges, provided the noise sequence $\{\boldsymbol{\eta}_k\}$ is diminishing and satisfies $\sum_{k=0}^{\infty} ||\boldsymbol{\eta}_k||_2<\infty.$
\end{restatable}

This result states that noise injection does not compromise the existing convergence properties of ADMM, if the noise amplitudes are chosen to satisfy an energy bound. The proof is provided in \suppref{apx:noise_inject}. We note that in addition to the theoretical guarantees we provided, our approach has a practical difference to noise injection in existing CM-based inverse problem solvers~\citep{garber2025CM4IR}, which uses a correction term based on the previous noise instance. Conversely, our approach generates the noise in a random manner, consistent with previous works from inverse problems~\citep{BM3Dicip2007,atchade2017perturbed}.

\looseness=-1
\paragraph{Momentum} To accelerate convergence and reduce the number of NFEs, we augment the ADMM updates with a momentum term. Momentum is commonly used in gradient descent type algorithms to improve convergence speed~\citep{nesterov1983method, thorley2021nesterov}. In the context of ADMM for imaging problems, these momentum updates have also been shown to reduce the number of iterations~\citep{goldstein2014fast,deng2025_SM_ADMM}, thus we adapt a similar scheme for PnP-CM.

\subsection{Final Algorithm} \label{sec:final_algo}
The complete algorithm is outlined in \algoref{alg:proposed_admm}. Our approach follows a PnP formulation in which CMs act as denoisers, together with noise perturbations and momentum-based updates to improve performance in the low-NFE regime while preserving convergence properties under suitable conditions. To remain consistent with common practices in diffusion and CM literature, we also adopt a reverse iteration ordering: unlike the standard ADMM algorithm, which counts iterations in ascending order, our algorithm enumerates iterations from a fixed number down to zero. 

\paragraph{Data fidelity update} The update in line~3 of \algoref{alg:proposed_admm} corresponds to a quadratic minimization problem, whose solution depends on the structure of the forward operator $\mathcal{A}$. For linear inverse problems with $\mathcal{A}(\x) = \mathbf{A}\x$, the update admits a closed-form solution:
\begin{equation}
\z_{n} = \big(\mathbf{A}^\top \mathbf{A} + \rho_{n+1} \mathbf{I}\big)^{-1} \big(\mathbf{A}^\top \mathbf{y} + \rho_{n+1} (\hat{\x}_{n+1} - \hat{\u}_{n+1})\big),
\end{equation}
which can be computed efficiently using matrix factorizations such as the singular value decomposition (SVD)~\citep{song2023pgdm,wang2023ddnm,garber2025CM4IR}, when applicable. For large-scale linear problems, such as MRI reconstruction, one can solve the system iteratively using conjugate gradient (CG) to avoid explicit matrix inversion, whose speed-up benefits over gradient descent-type approaches has also been noted in the context of DM-based inverse problem solvers~\citep{chung2024decomposed}. Finally, for more general nonlinear forward models, the subproblem can be solved using standard first-order optimization methods, such as gradient descent (GD) or Adam. Further implementation details, including coefficient choices, are provided in \suppref{appdx:details}. Additionally, sensitivity analysis for hyperparameters is provided in \suppref{appdx:sensitivity}.

\input{Tables/table_quantitative}
\section{Experiments}
\subsection{Experimental Setup and Model Implementation Details} \label{sec:exp}
We conduct comprehensive evaluations on the LSUN Bedroom~\citep{yu2015lsun} and CelebA-HQ~\citep{karras2018progressive} datasets at an image resolution of $256 \times 256 \times 3$, and on the fastMRI~\citep{knoll2020fastmri_dataset-journal} dataset with image size of $320 \times 320 \times 2$. In particular, the New York University fastMRI dataset~\citep{knoll2020fastmri_dataset-journal} contains complex-valued images and multi-coil k-space data from fully sampled coronal proton density (PD) and proton density with fat suppression (PD-FS) knee MRI scans, obtained with relevant institutional review board approvals.

\begin{figure*}[t]
    \centering
    \includegraphics[width=1.0\linewidth]{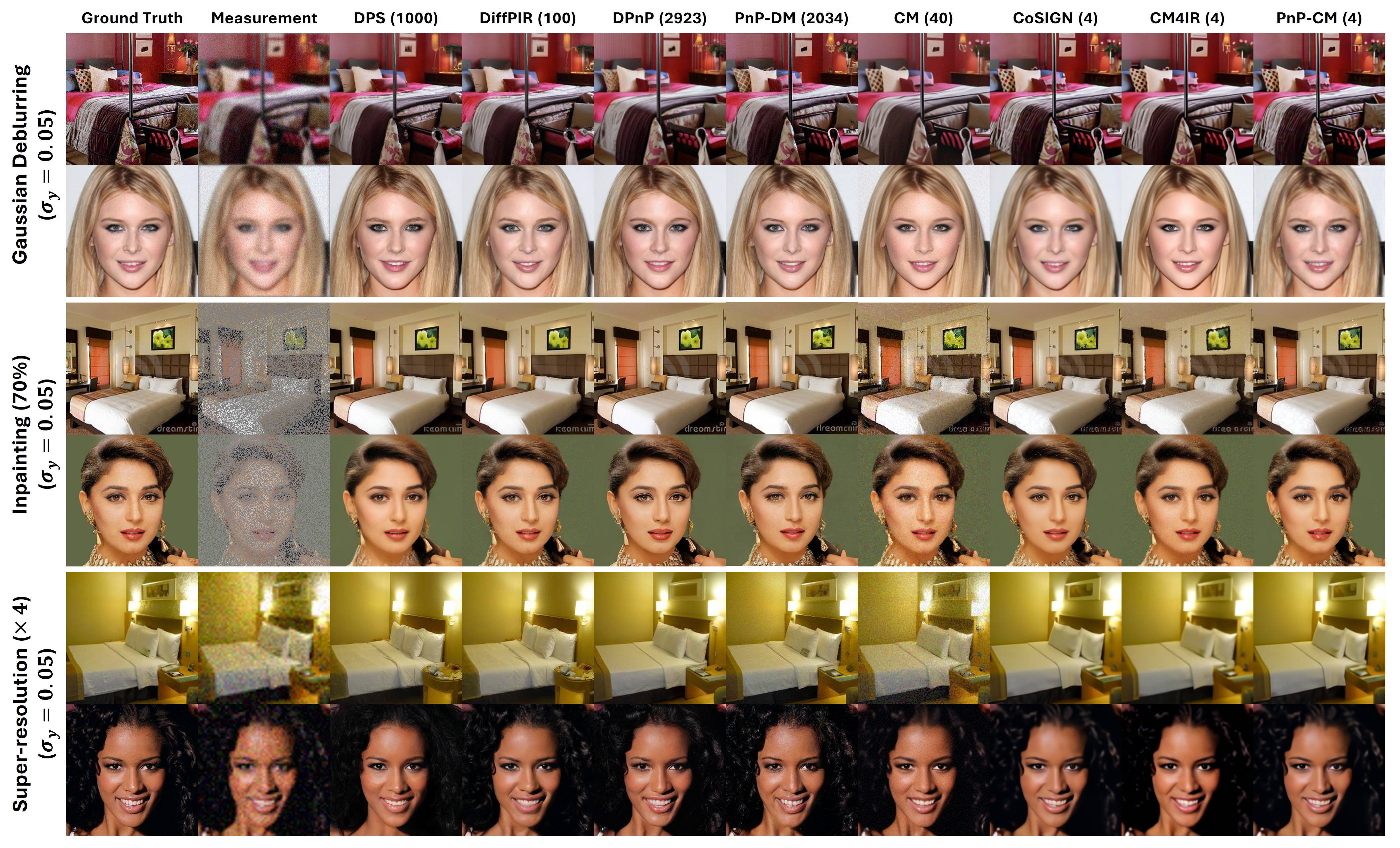} 
    \caption{Representative results for Gaussian deblurring, inpainting (70\%), and super-resolution (×4). PnP-CM produces sharp and coherent reconstructions, preserving fine details while avoiding the over-smoothing observed in DPS. Compared to CM-based methods, it more reliably recovers structured content, and achieves visual quality comparable to recent DM-based PnP methods (\eg, DiffPIR, DPnP, PnP-DM) while requiring substantially fewer iterations.}
    \vspace{-1ex}
    \label{fig:main_inv_all}
\end{figure*}

For LSUN Bedroom, we use the pre-trained unconditional CM provided by~\citep{song2023_consistency_models} without additional retraining. For both CelebA-HQ and MRI datasets, we train a CM from scratch by first training an EDM model~\citep{karras2022elucidating}. For CelebA-HQ, we use 27k training images as in~\citep{kingma2018glow}. For MRI, we use all 973 volumes (subjects) from the training set of the fastMRI dataset, excluding the first and last five slices of each volume, following the prior protocols~\citep{chung2022scoreMRI, chung2024decomposed}. We provide further details for training the EDM and CM networks in \suppref{apx:DM_CM_4_MRI}.

For evaluation on natural image tasks, we sample 300 images from the LSUN Bedroom and CelebA-HQ validation sets, respectively. For evaluation on MRI reconstruction, we apply the aforementioned slice exclusion protocol on 10 validation volumes from fastMRI, yielding 228 slices for PD and 240 slices for PD-FS contrasts.

\subsection{Experiments on Inverse Problems}
\paragraph{Problem setup} We evaluate our PnP-CM algorithm on a set of noisy inverse problems. The following tasks are used on the LSUN bedroom and CelebA-HQ datasets: For \emph{linear} operators, we consider (i) random inpainting with 70\% masking, (ii) bicubic super-resolution with 4$\times$ downsampling, and (iii) Gaussian deblurring using a $5\times5$ separable kernel with standard deviation $\sigma = 10.0$. For \emph{non-linear} operators, the tasks include (i) JPEG compression with quality factor (QF) 5, (ii) nonlinear deblurring with a neural network–approximated forward model~\citep{tran2021explore}, and (iii) phase retrieval following the protocol in~\citep{xu2024_DPnP} with a coded mask~\citep{candes2015phase}. All measurements are generated by applying the forward operator to the ground-truth images, with additive Gaussian measurement noise level of $\sigma_y=0.05$. 

For a practical real-world application, multi-coil MRI reconstruction is considered, which is an inherently noisy linear inverse problem. Multi-coil complex-valued raw MRI datasets are retrospectively undersampled from \emph{noisy} fully-sampled raw MRI k-space measurements using 1D random subsampling. Acceleration factors, $R$ of 4 and 8 are used with 24 and 12 central k-space lines, respectively, as in~\citep{chung2024decomposed}.

\begin{figure}[!t]
    \centering
    \includegraphics[width=1.0\linewidth]{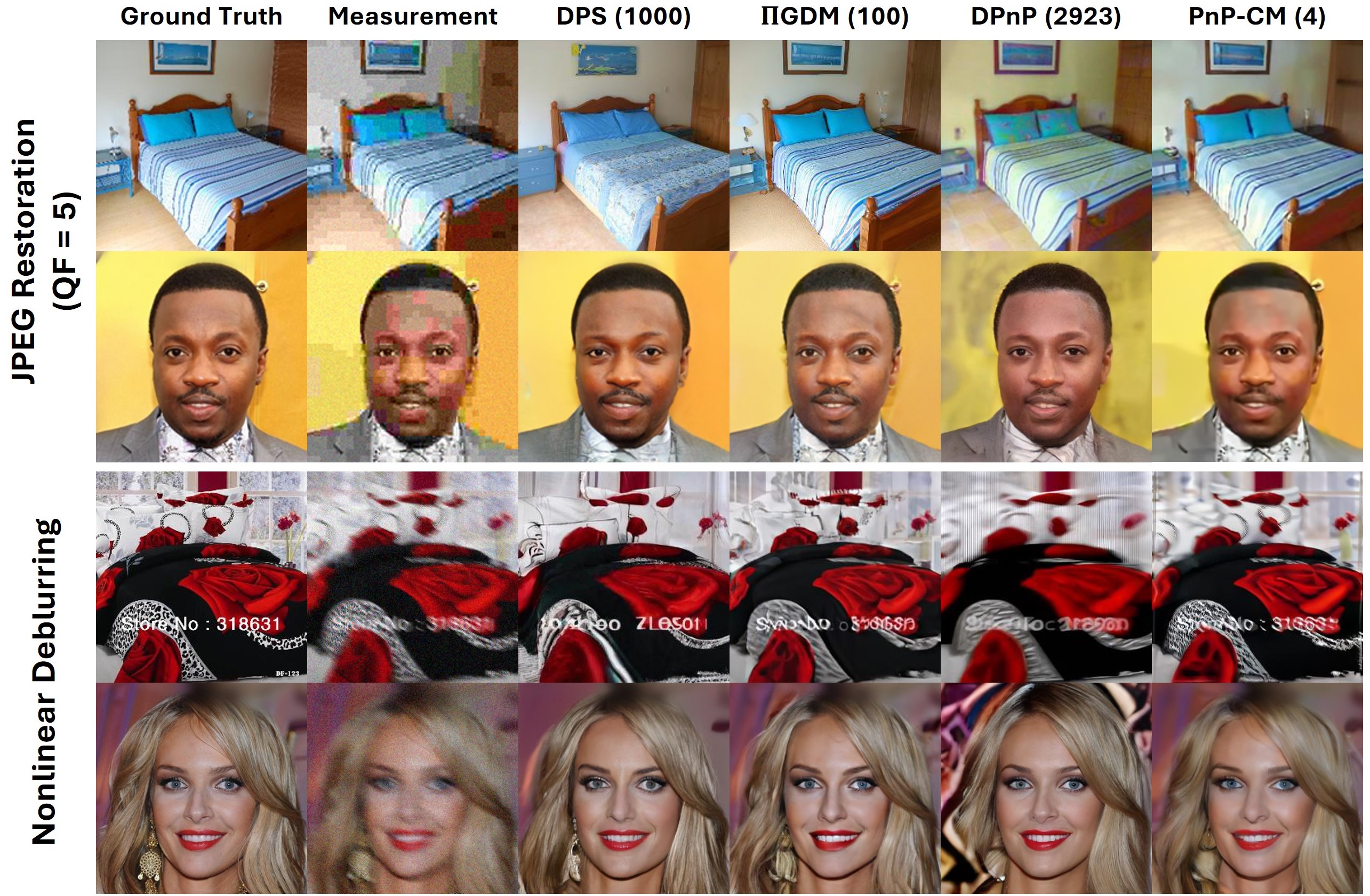}\\
    \vspace{1ex}
    \includegraphics[width=1.0\linewidth]{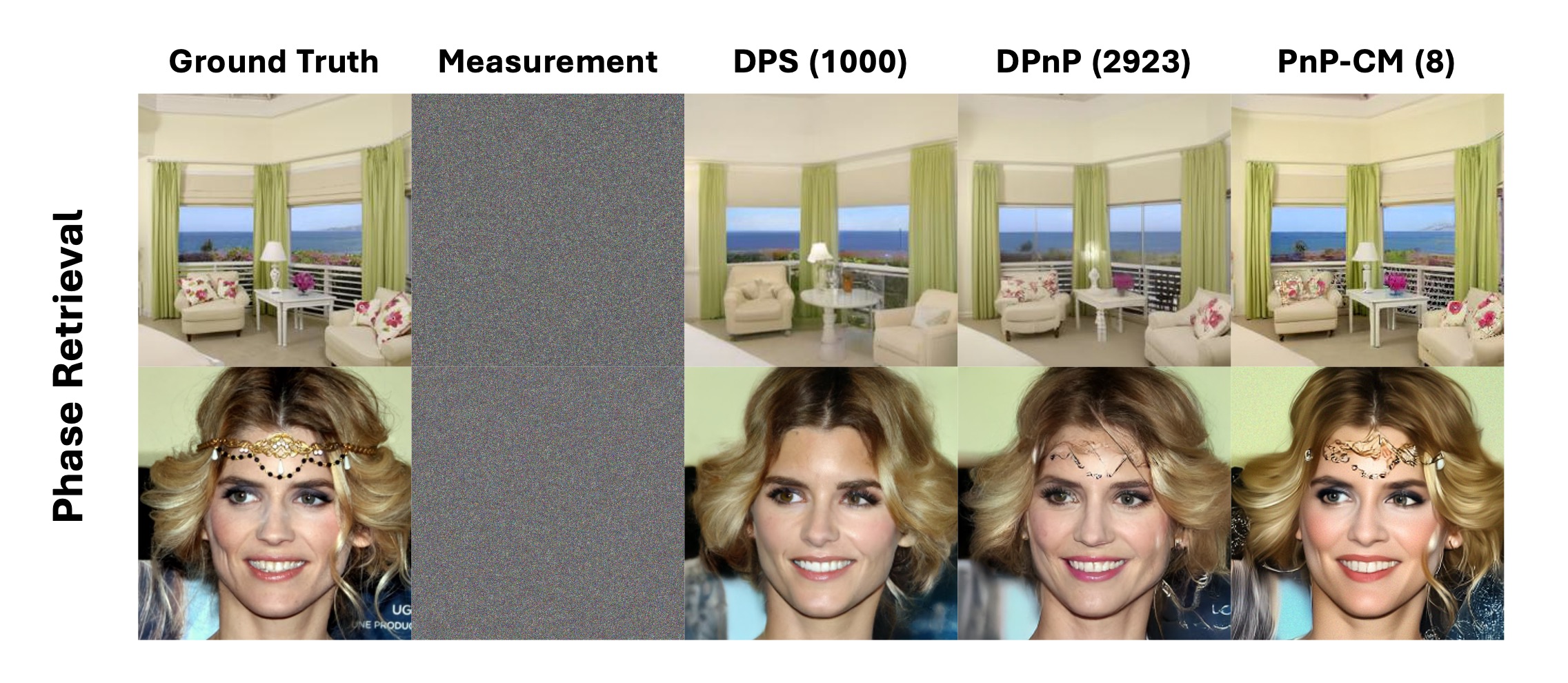}
    \vspace{-3ex}
    \caption{Comparison of reconstruction quality for nonlinear forward models. PnP-CM yields sharp and coherent reconstructions, with quality that is competitive with or improved over existing methods, highlighting robust performance in challenging nonlinear settings while requiring substantially fewer NFEs.} 
    \label{fig:nonlinear_main}
    \vspace{-3ex}
\end{figure}

\paragraph{Comparison methods} We conducted extensive comparisons with various methods. For experiments on natural images, we compare our method with DPS~\citep{chung2023dps}, $\Pi$GDM~\citep{song2023pgdm}, DiffPIR~\citep{zhu2023DiffPIR}, DPnP~\citep{xu2024_DPnP} and PnP-DM~\citep{wu2024_PnP-DM} as DM-based approaches, and CM~\citep{song2023_consistency_models}, CoSIGN~\citep{zhao2024cosign}, and CM4IR~\citep{garber2025CM4IR} as CM-based solvers. For MRI reconstruction, we benchmark against DPS and DDS~\citep{chung2024decomposed}, both of which employ DM priors, as well as CM4IR. To ensure a fair comparison, we employed the same unconditional diffusion and consistency models across all algorithms. For DM-based methods, we used the pre-trained unconditional DM on LSUN Bedroom from~\citep{dhariwal2021beatGANs}, as well as models pre-trained on CelebA-HQ from~\citep{choi2021ILVR} and fastMRI from~\citep{chung2024decomposed}. Further implementation details are provided in \suppref{appdx:comparison}.

\begin{figure}[!b]
    \vspace{-1.5ex}
    \centering
    \includegraphics[width=1.0\linewidth]{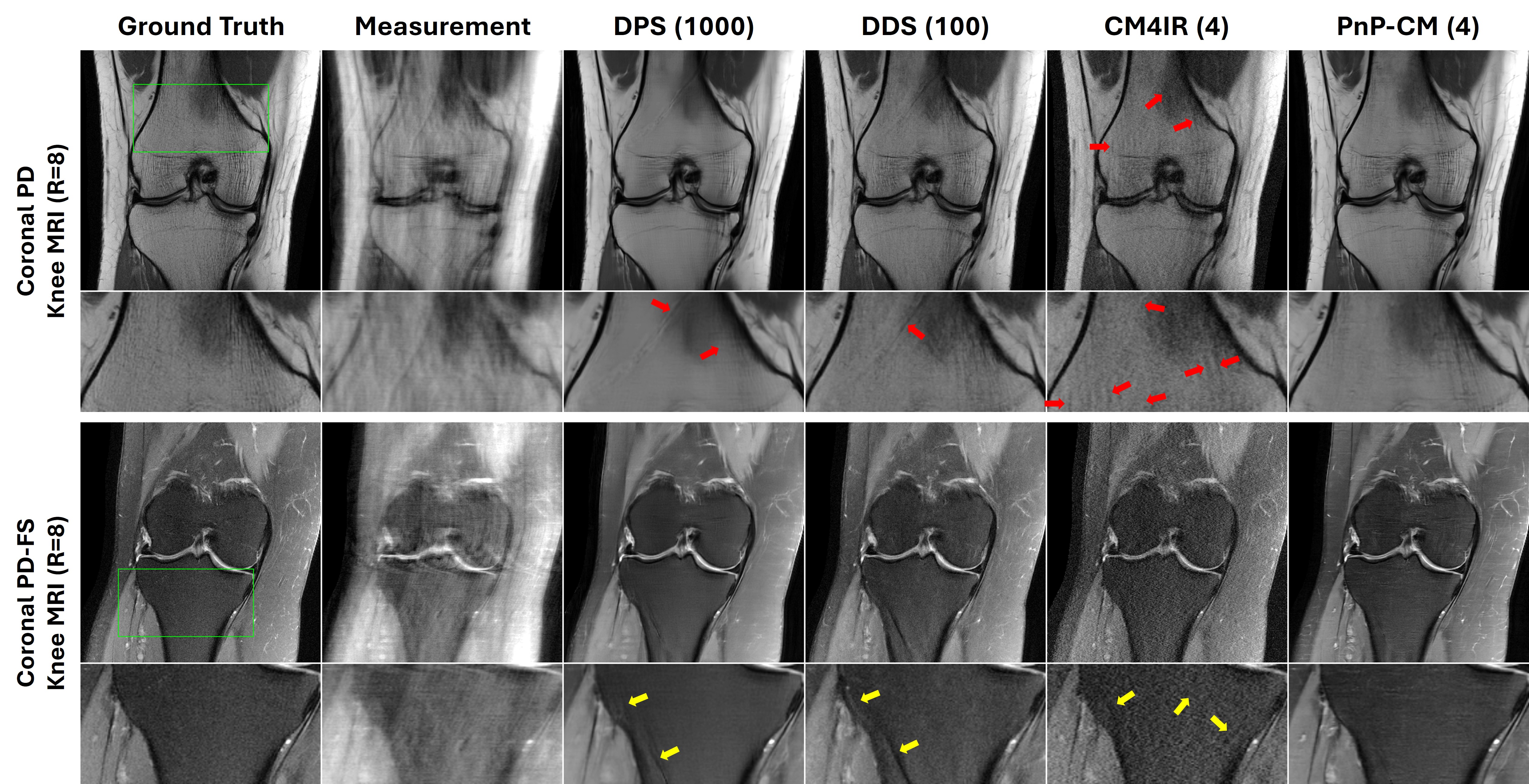} 
     \caption{Qualitative comparisons of DPS, DDS, CM4IR, and PnP-CM. \textbf{Top:} Coronal PD with $R=8$. \textbf{Bottom:} Coronal PD-FS with $R=8$. 
    PnP-CM effectively reduces artifacts and blurring that are not removed by other methods (red and yellow arrows).}
    \label{fig:MRI_R8_main}
\end{figure}

\subsection{Quantitative and Qualitative Results}
Given the differing requirements of natural and medical image tasks, we employ separate evaluation metrics for the two settings. For natural image tasks, reconstruction quality is assessed using peak signal-to-noise-ratio (PSNR) and learned perceptual image patch similarity (LPIPS), while for medical imaging tasks, we use PSNR and structural similarity index (SSIM) as evaluation metrics. Quantitative results are reported as averages over the validation set, while qualitative examples are provided to illustrate the visual fidelity of the reconstructions. For all tasks, we use $\mathrm{N}=4$ iterations, except for phase retrieval where we use $\mathrm{N}=8$ due to its increased difficulty. In \suppref{appdx:diff_nfe_and_noise}, we further report results for $\mathrm{N}\in\{2,4,8\}$ and $\sigma_\y=0.025$ to illustrate the performance and robustness of the method.

\paragraph{Natural image tasks} Representative results for linear inverse problems in \figref{fig:main_inv_all} demonstrate that PnP-CM consistently yields sharp and high-quality reconstructions. Compared to DPS, our method avoids oversmoothing while being significantly faster, and it generates coherent structures that CM-based approaches often fail to recover. The reconstructions also preserve high frequency details without introducing the grid-like artifacts seen in some of the competing methods. Compared to recent DM-based PnP methods, which can achieve high-quality reconstructions but typically require substantially more iterations, PnP-CM attains comparable visual quality while operating at a fraction of the computational cost. For nonlinear inverse problems (\figref{fig:nonlinear_main}), which are intrinsically difficult to solve, PnP-CM yields reconstructions that are either improved or comparable to those of DPS, $\Pi$GDM, and DPnP, while requiring significantly fewer NFEs. These visual observations are reflected quantitatively in \tabref{tab:comparison_method}, where PnP-CM reaches state of the art quality with only a few evaluations. Additional qualitative comparisons are provided in \suppref{appdx:celeba_appdx}.

\paragraph{MRI reconstruction} 
Representative MRI reconstructions for $R=8$ are depicted in \figref{fig:MRI_R8_main}, showing that our method substantially reduces blurring artifacts observed in DPS, as well as structured artifacts present in both DPS and DDS. CM4IR exhibits residual aliasing and noise amplification in this scenario, as the back-projection term in its data fidelity, ${\bf A}^\dagger({\bf Ax - y})$ amounts to ${\bf x} - {\bf A}^\dagger {\bf y}$ in multi-coil MRI, where $m > n$, and ${\bf A}^\dagger = ({\bf A}^\top {\bf A})^{-1} {\bf A}^\top$. In other words, the solution moves toward the linear least squares solution, ${\bf A}^\dagger {\bf y}$, which suffers from substantial aliasing artifacts. \tabref{tab:comparison_method} summarizes the performance of different approaches on the Coronal PD and PD-FS datasets with acceleration rates of $R=4$ and $R=8$. Across all settings, our proposed method with $\text{NFE}=4$ consistently outperforms DPS (NFE=1000), DDS (NFE=100) and CM4IR (NFE=4) in terms of PSNR and SSIM. Additional visual results, including those at the more modest $R=4$, are provided in \suppref{apx:MRI_recon}.

\input{Tables/table_ablation_celeba}

\subsection{Ablation Studies}
We conducted two ablation studies to disentangle the contributions of momentum and noise injection in our ADMM-based solver. The first ablation study compared four variants: (i) PnP-CM without momentum and noise injection, (ii) PnP-CM without momentum, (iii) PnP-CM without noise injection, and (iv) our proposed PnP-CM. Experiments were performed on the CelebA-HQ dataset for super-resolution, inpainting, and Gaussian deblurring with $\text{NFE}=4$ and $\sigma_y=0.05$. The quantitative results given in \tabref{tab:ADMM_method_comparison} consistently demonstrate that both momentum and noise injection improve reconstruction quality. Our second ablation study, provided in \suppref{appdx:second_ablation} examines the effect of momentum across varying NFEs, and shows momentum improves performance primarily in the low-NFE regime with diminishing gains at higher NFEs.

\section{Conclusions}
In this work, we introduce PnP-CM, a novel framework that integrates consistency models (CMs) as proximal operators within the PnP optimization paradigm. By combining the efficiency of CMs with principled optimization-based updates, PnP-CM offers a significant advance in solving inverse problems. Our method, enhanced with noise injection and momentum, achieves high-quality reconstructions in as few as 2-4 NFEs, demonstrating both precision and speed. We evaluate PnP-CM across diverse linear and nonlinear inverse problems, including natural image restoration and medical imaging tasks. We also show applications in MRI reconstruction, where we demonstrate the \emph{first} application of CMs trained specifically for MRI data. Through extensive evaluations, we demonstrate that PnP-CM surpasses existing state-of-the-art CM-based methods in reconstruction quality, setting a new benchmark for fast and high-fidelity inverse problem solving.

{
    \small
    \bibliographystyle{misc/ieeenat_fullname}
    \bibliography{refs}
}

\input{supp_mat}

\end{document}

%% file: misc/math_commands.tex
\usepackage{amsmath,amsfonts,bm}

\def\figref#1{figure~\ref{#1}}
\def\secref#1{section~\ref{#1}}
\def\eqref#1{equation~\ref{#1}}
\def\1{\bm{1}}

\DeclareMathAlphabet{\mathsfit}{\encodingdefault}{\sfdefault}{m}{sl}
\SetMathAlphabet{\mathsfit}{bold}{\encodingdefault}{\sfdefault}{bx}{n}

%% file: misc/preamble.tex
\usepackage{makecell}
\definecolor{junno_purple}{RGB}{150, 0, 200}
\definecolor{utku_green}{rgb}{0.0, 0.6, 0.0}
\definecolor{merve_blue}{rgb}{0.1, 0.2, 0.6}
\usepackage[]{color-edits}
\addauthor{ma}{red}
\addauthor{mg}{merve_blue}
\addauthor{ua}{utku_green}
\addauthor{jy}{junno_purple}
\definecolor{gray_utku}{rgb}{0.3, 0.3, 0.3}
\newcommand{\var}[1]{\textcolor{gray_utku}{\text{\tiny$\pm$#1}}}

 \renewcommand{\paragraph}[1]{\vspace{.5em}\noindent\textbf{#1.}}

\makeatletter
\usepackage{xspace}
\def\@onedot{\ifx\@let@token.\else.\null\fi\xspace}
\DeclareRobustCommand\onedot{\futurelet\@let@token\@onedot}
\newcommand{\eqnref}[1]{Eq\onedot~\ref{#1}}
\renewcommand{\figref}[1]{Fig\onedot~\ref{#1}}
\newcommand{\algoref}[1]{Alg\onedot~\ref{#1}}
\renewcommand{\secref}[1]{Section~\ref{#1}}
\newcommand{\tabref}[1]{Tab\onedot~\ref{#1}}
\newcommand{\suppref}[1]{SuppMat~\ref{#1}}

\def\eg{\emph{e.g}\onedot}
\def\ie{\emph{i.e}\onedot}
\def\iid{i.i.d\onedot}

%% file: Algos/pnp_cm_algo.tex
\begin{algorithm*}[t]
\setstretch{1.12}
\caption{PnP-CM for General Inverse Problems}
\label{alg:proposed_admm}
\begin{algorithmic}[1]
    \Statex \hspace{-1.5em} \textbf{Require:} Consistency model $f_\theta(\cdot,t)$, sequence of time points $\{t_n\}_{n=0}^{\mathrm{N}}$, CM noise level scale $\{\delta_n\}_{n=1}^{\mathrm{N}} > 0$, 
    \Statex \hspace*{2.6em} penalty parameters $\{\rho_n\}_{n=1}^{\mathrm{N}} > 0$, momentum coefficients $\{\mu_n\}_{n=1}^{\mathrm{N}} > 0$, 
    forward operator $\mathcal{A}$, measurement $\mathbf{y}$
    \State \textbf{Initialize:} $\hat{\x}_{\mathrm{N}}\gets 0,\;\hat{\u}_{\mathrm{N}}\gets 0,\;\x_{\mathrm{N}}\gets 0,\;\u_{\mathrm{N}}\gets 0$ 
    \For {$n$ from $\mathrm{N}-1$ to $0$}
        \vspace{1mm}
        \State $\z_{n} = \displaystyle \arg\min_{\z} \tfrac{1}{2}\|\mathcal{A}(\z) - \mathbf{y}\|_2^2 + \tfrac{\rho_{n+1}}{2}\|\z - (\hat{\x}_{n+1} - \hat{\u}_{n+1})\|_2^2$
        \hfill {\color{gray}\(\triangleright\) Data-fidelity update}
        \vspace{1mm}
        \State $\boldsymbol{\nu}_{n} \sim \mathcal{N}\!\big(\z_{n}+\hat{\u}_{n+1},t_{n+1}^2\mathbf{I} \big)$
        \hfill {\color{gray}\(\triangleright\) Noise injection}
        \vspace{1mm}
        \State $\x_{n} = f_\theta\big(\boldsymbol{\nu}_{n}, (1+\delta_{n+1}) \, t_{n+1} \big)$
        \hfill {\color{gray}\(\triangleright\) CM proximal operator}
        \vspace{1mm}
        \State $\u_{n} = \hat{\u}_{n+1} + \z_{n} - \x_{n}$
        \hfill {\color{gray}\(\triangleright\) Dual variable update}
        \vspace{1mm}
        \State $\hat{\x}_{n} = \x_{n} + \mu_{n+1}(\x_{n} - \x_{n+1})$
        \hfill {\color{gray}\(\triangleright\) Primal momentum update}
        \vspace{1mm}
        \State $\hat{\u}_{n} = \u_{n} + \mu_{n+1}(\u_{n} - \u_{n+1})$
        \hfill {\color{gray}\(\triangleright\) Dual momentum update}
        \vspace{1mm}
    \EndFor
    \State \textbf{Return:} $\x_0$
\end{algorithmic}
\end{algorithm*}

%% file: Tables/table_quantitative.tex
\begin{table*}[t]
\caption{Quantitative comparison of natural-image and MRI inverse problems. For natural images we use $\sigma_y=0.05$, while MRI reconstructions contain inherent measurement noise. Best: \textbf{BOLD}, second-best: \underline{Underlined}. Methods are omitted if they cannot be reliably implemented for the given task.}
\vspace{-4mm}
\label{tab:comparison_method}
\begin{center}
\begin{small}
\renewcommand{\arraystretch}{1.0}
\setlength{\tabcolsep}{6.5pt}

\begin{tabular}{lccccccc}
\toprule

% ============================================================
% LINEAR NATURAL IMAGING SECTION HEADER
% ============================================================
\rowcolor{gray!20}
\multicolumn{8}{c}{\textbf{Linear Inverse Problems on Natural Images (CelebA-HQ~\cite{karras2018progressive} and LSUN Bedroom~\cite{yu2015lsun})}} \\
\midrule

\multirow{2}{*}{\textbf{Method}} & \multirow{2}{*}{\textbf{NFE$\downarrow$}} &
\multicolumn{3}{c}{\textbf{CelebA-HQ (PSNR$\uparrow$ / LPIPS$\downarrow$)}} &
\multicolumn{3}{c}{\textbf{LSUN Bedroom (PSNR$\uparrow$ / LPIPS$\downarrow$)}} \\
\cmidrule(lr){3-5} \cmidrule(lr){6-8}
& &
{\small SR $\times$4} &
{\small Gauss. Deblur} &
{\small Inpainting} &
{\small SR $\times$4} &
{\small Gauss. Deblur} &
{\small Inpainting} \\
\midrule

\addlinespace[2pt]
DPS~\citep{chung2023dps} & 1000 
& 24.48 / \underline{0.253} & 25.36 / \underline{0.225} & 28.33 / \textbf{0.194} 
& 22.19 / 0.335 & 22.88 / 0.286 & 26.38 / 0.231 \\
\arrayrulecolor{gray!40}
\addlinespace[1pt]\midrule
\addlinespace[2pt]

\addlinespace[1pt]
DiffPIR~\citep{zhu2023DiffPIR} & 100
& 26.30 / 0.271 & \underline{27.32} / 0.246 & \textbf{30.71} / 0.207
& 23.78 / 0.334 & 24.60 / 0.303 & \textbf{28.66} / \underline{0.212} \\
\arrayrulecolor{gray!40}
\addlinespace[1pt]\midrule
\addlinespace[2pt]

\addlinespace[1pt]
DPnP~\citep{xu2024_DPnP} & 2923 
& 26.21 / 0.259 & 26.48 / 0.252 & 28.33 / 0.222
& 23.79 / \underline{0.317} & 24.11 / 0.303 & 27.10 / 0.230 \\
\arrayrulecolor{gray!40}
\addlinespace[1pt]\midrule
\addlinespace[2pt]

PnP-DM~\citep{wu2024_PnP-DM} & 2034
& 26.70 / \textbf{0.230} & 26.99 / \textbf{0.218} & 27.80 / \underline{0.195}
& 24.43 / \textbf{0.270} & 24.66 / \textbf{0.257} & 26.40 / \textbf{0.209} \\
\addlinespace[2pt]
\arrayrulecolor{orange!60} \hdashline

\addlinespace[1pt]
CM~\citep{song2023_consistency_models} & 40 
& 25.77 / 0.431 & 26.81 / 0.381 & 27.02 / 0.344
& 23.16 / 0.425 & 24.86 / 0.388 & 24.44 / 0.410 \\
\arrayrulecolor{gray!40}
\addlinespace[1pt]\midrule

\addlinespace[1pt]
CoSIGN~\citep{zhao2024cosign}
& 4 
& 22.08 / 0.594 & 22.87 / 0.496 & 27.03 / 0.367 
& 20.90 / 0.545 & 23.08 / 0.441 & 25.68 / 0.339  \\
\arrayrulecolor{gray!40}
\addlinespace[1pt]\midrule

\addlinespace[1pt]
CM4IR~\citep{garber2025CM4IR}

& 4 
& \underline{26.77} / 0.392 & 27.30 / 0.297 & 28.21 / 0.345
& \underline{25.50} / 0.334 & \underline{27.37} / \underline{0.270} & 26.78 / 0.303 \\
\arrayrulecolor{gray!40}
\addlinespace[1pt]\midrule

\addlinespace[1pt]

PnP-CM \textbf{(ours)}

& 4
& \cellcolor{brown!30} \textbf{27.27} / 0.285 
& \cellcolor{brown!30} \textbf{28.94} / 0.249 
& \cellcolor{brown!30} \underline{29.23} / 0.201 
& \cellcolor{brown!30} \textbf{25.56} / 0.321 
& \cellcolor{brown!30} \textbf{27.49} / 0.285 
& \cellcolor{brown!30} \underline{27.12} / 0.233 \\
\arrayrulecolor{black} \midrule

% ============================================================
% NONLINEAR NATURAL IMAGING SECTION HEADER
% ============================================================
\rowcolor{gray!20}
\multicolumn{8}{c}{\textbf{Nonlinear Inverse Problems on Natural Images (CelebA-HQ~\cite{karras2018progressive} and LSUN Bedroom~\cite{yu2015lsun})}} \\
\midrule

\multirow{2}{*}{\textbf{Method}} & \multirow{2}{*}{\textbf{NFE$\downarrow$}} &
\multicolumn{3}{c}{\textbf{CelebA-HQ (PSNR$\uparrow$ / LPIPS$\downarrow$)}} &
\multicolumn{3}{c}{\textbf{LSUN Bedroom (PSNR$\uparrow$ / LPIPS$\downarrow$)}} \\
\cmidrule(lr){3-5} \cmidrule(lr){6-8}
& &
{\small JPEG Res.} &
{\small Nonlin. Deblur} &
{\small Phase Ret.} &
{\small JPEG Res.} &
{\small Nonlin. Deblur} &
{\small Phase Ret.} \\
\midrule

\addlinespace[1pt]
DPS~\citep{chung2023dps} & 1000 
& \underline{25.55} / \underline{0.234} & 23.55 / \underline{0.253} & 21.38 / \textbf{0.327} & 20.23 / \underline{0.397} & 21.19 / \underline{0.324} & 18.06 / \underline{0.417} \\
\arrayrulecolor{gray!40}
\addlinespace[1pt]\midrule
\addlinespace[2pt]

\addlinespace[1pt]
$\Pi$GDM~\citep{song2023pgdm} & 100
& 25.23 / \textbf{0.221} & \underline{24.13} / \textbf{0.235} & \multicolumn{1}{c}{--}
& \underline{23.34} / \textbf{0.285} & \underline{22.52} / \textbf{0.290} & \multicolumn{1}{c}{--} \\
\arrayrulecolor{gray!40}
\addlinespace[1pt]\midrule
\addlinespace[2pt]

DPnP~\citep{xu2024_DPnP} & 2923 
& 21.31 / 0.402 & 22.29 / 0.333 & \textbf{22.89} / \underline{0.338} & 20.24 / 0.510  & 21.06 / 0.391 & \underline{18.55} / \textbf{0.389} \\

\addlinespace[2pt]
\arrayrulecolor{orange!60} \hdashline

\addlinespace[2pt]
PnP-CM \textbf{(ours)}
& 4 or 8 
& \cellcolor{brown!30} \textbf{26.14} / 0.356 
& \cellcolor{brown!30} \textbf{24.55} / 0.304 
& \cellcolor{brown!30} \underline{22.18} / 0.422
& \cellcolor{brown!30} \textbf{23.49} / 0.410 
& \cellcolor{brown!30} \textbf{22.54} / 0.381
& \cellcolor{brown!30} \textbf{18.98} / 0.523 \\
\arrayrulecolor{black}\midrule

% ============================================================
% MRI SECTION HEADER
% ============================================================
\rowcolor{gray!20}
\multicolumn{8}{c}{\textbf{Medical Imaging (MRI Reconstruction using fastMRI~\cite{knoll2020fastmri_dataset-journal})}} \\
\midrule

\multirow{2}{*}{\textbf{Method}} & \multirow{2}{*}{\textbf{NFE$\downarrow$}} & \multirow{2}{*}{\textbf{CG iter.$\downarrow$}} & \multirow{2}{*}{\textbf{R}} &
\multicolumn{2}{c}{\textbf{Coronal PD}} &
\multicolumn{2}{c}{\textbf{Coronal PD-FS}} \\
\cmidrule(lr){5-6} \cmidrule(lr){7-8}
& & & &
{\small PSNR$\uparrow$} & {\small SSIM$\uparrow$} &
{\small PSNR$\uparrow$} & {\small SSIM$\uparrow$} \\
\midrule

\addlinespace[-1pt]
\multirow{2}{*}{DPS~\citep{chung2023dps}} 
& \multirow{2}{*}{1000} & \multirow{2}{*}{--}
& {\scriptsize $\times 4$}
& 34.00\var{2.53} & 0.893\var{0.036}
& \underline{31.68}\var{3.21} & \underline{0.796}\var{0.075} \\
& & 
& {\scriptsize $\times 8$}
& 31.88\var{2.89} & 0.863\var{0.041}
& \underline{29.55}\var{3.71} & \underline{0.758}\var{0.085} \\
\arrayrulecolor{gray!40}
\addlinespace[-1pt]\midrule

\addlinespace[-1pt]
\multirow{2}{*}{DDS~\citep{chung2024decomposed}} 
& \multirow{2}{*}{100} & \multirow{2}{*}{10}
& {\scriptsize $\times 4$}
& \underline{34.39}\var{1.65} & \underline{0.907}\var{0.029}
& 31.14\var{2.26} & 0.792\var{0.066} \\
& & 
& {\scriptsize $\times 8$}
& \underline{32.27}\var{1.68} & \underline{0.868}\var{0.038}
& 29.18\var{2.53} & 0.740\var{0.082} \\

\addlinespace[2pt]
\arrayrulecolor{orange!60} \hdashline 
\addlinespace[1pt]

\multirow{2}{*}{CM4IR~\citep{garber2025CM4IR}} 
& \multirow{2}{*}{4} & \multirow{2}{*}{--}
& {\scriptsize $\times 4$}
& 31.05\var{2.01} & 0.796\var{0.067}
& 26.51\var{2.77} & 0.600\var{0.120} \\
& & 
& {\scriptsize $\times 8$}
& 29.60\var{1.76} & 0.755\var{0.071}
& 25.36\var{2.75} & 0.556\var{0.126} \\
\arrayrulecolor{gray!40}
\addlinespace[-1pt]\midrule

\addlinespace[-1pt]
\multirow{2}{*}{PnP-CM \textbf{(ours)}} 
& \multirow{2}{*}{4} & \multirow{2}{*}{10}
& {\scriptsize $\times 4$}
& \cellcolor{brown!30}\textbf{35.57}\var{1.49} & \cellcolor{brown!30}\textbf{0.915}\var{0.026}
& \cellcolor{brown!30}\textbf{32.90}\var{2.32} & \cellcolor{brown!30}\textbf{0.842}\var{0.050} \\
& & 
& {\scriptsize $\times 8$}
& \cellcolor{brown!30}\textbf{33.24}\var{1.69} & \cellcolor{brown!30}\textbf{0.884}\var{0.033}
& \cellcolor{brown!30}\textbf{31.42}\var{2.20} & \cellcolor{brown!30}\textbf{0.804}\var{0.060} \\
\arrayrulecolor{black}
\addlinespace[-1pt]\bottomrule

\end{tabular}
\end{small}
\end{center}
\vspace{-3ex}
\end{table*}

%% file: Tables/table_ablation_celeba.tex
\begin{table}[!t]%[15]{r}{0.65\textwidth}
\vspace{1.5ex}
\captionsetup{type=table}
\caption{Comparison of ADMM strategies and highlighting the effect of momentum and noise injection ($\sigma_{y}=0.05$, $\textrm{N}=4$). Best: {\bf BOLD}, second-best: \underline{Underlined}.}
\vspace{-3.5mm}
\label{tab:ADMM_method_comparison}
\begin{center}
\begin{footnotesize}
\renewcommand{\arraystretch}{1.1}
\setlength{\tabcolsep}{2pt}
\begin{tabular}{ccccc}
\toprule
\multicolumn{1}{c}{\textbf{Noise Inj.}} &
\multicolumn{1}{c}{\textbf{Moment.}} &
\multicolumn{1}{c}{\textbf{SR $\times$4}} &
\multicolumn{1}{c}{\textbf{\shortstack{Gaussian \\ Deblurring}}} &
\multicolumn{1}{c}{\textbf{Inpainting}} \\
\cmidrule(lr){3-5}
  & & {\scriptsize PSNR$\uparrow$} / {\scriptsize LPIPS$\downarrow$} &
      {\scriptsize PSNR$\uparrow$} / {\scriptsize LPIPS$\downarrow$} &
      {\scriptsize PSNR$\uparrow$} / {\scriptsize LPIPS$\downarrow$} \\
\midrule

\xmark & \xmark  
    & 22.93 / 0.461 
    & 27.26 / 0.427 
    & 28.69 / 0.295 \\
\arrayrulecolor{gray!40}\midrule

\cmark & \xmark
    & 24.25 / \underline{0.300} 
    & \underline{28.89} / \underline{0.251} 
    & \underline{29.13} / \underline{0.204} \\
\arrayrulecolor{gray!40}\midrule

\xmark & \cmark  
    & \underline{25.97} / 0.469 
    & 26.59 / 0.453 
    & 28.72 / 0.293 \\
\arrayrulecolor{gray!40}\midrule

\cmark & \cmark
    & \cellcolor{brown!30} \textbf{27.27} / \textbf{0.285} 
    & \cellcolor{brown!30} \textbf{28.94} / \textbf{0.249} 
    & \cellcolor{brown!30} \textbf{29.23} / \textbf{0.201} \\
\arrayrulecolor{black}\bottomrule
\end{tabular}
\end{footnotesize}
\end{center}
\vspace{-3ex}
\end{table}

%% file: supp_mat.tex
\clearpage
\setcounter{page}{12}
\maketitlesupplementary

\appendix

\section{Proof of Theorem 1}
\label{apx:noise_inject}
\begin{proof}
To analyze the effect of noise injection on the convergence behavior, we compare the combined residuals in the presence and absence of noise, decomposing the residual into its primal and dual components:
{\small
\begin{equation}
    \Delta_k
    =\frac{1}{\sqrt{P}}\Big(
    \underbrace{||\z_{k+1}-\z_k||_2}_{\Delta_{\z_k}}
    + \underbrace{||\x_{k+1}-\x_k||_2}_{\Delta_{\x_k}}
    + \underbrace{||\u_{k+1}-\u_k||_2}_{\Delta_{\u_k}}
    \Big).
\end{equation}
}
Since the primal variable $\z$ is updated prior to noise injection, it remains unaffected. We use the superscripts $\boldsymbol{0}$ and $\boldsymbol{\eta}$ to denote the cases without and with noise injection, respectively.

For the $\x$-update, we obtain
\begin{align}
    \Delta_{\x_k}^{\boldsymbol{\eta}}
    &= ||\x_{k+1}-\x_k||_2 \nonumber \\
    &= ||f_\theta(\z_{k+1}+\u_{k}+\boldsymbol{\eta}_k)-\x_k||_2 \nonumber \\
    &= ||f_\theta(\z_{k+1}+\u_{k}+\boldsymbol{\eta}_k) 
    - f_\theta(\z_{k+1}+\u_{k}) \notag\\
    &\quad + f_\theta(\z_{k+1}+\u_{k}) 
    -\x_k||_2 \nonumber \\
    &\leq ||f_\theta(\z_{k+1}+\u_{k}+\boldsymbol{\eta}_k) 
    - f_\theta(\z_{k+1}+\u_{k})||_2 \notag\\
    &\quad + ||f_\theta(\z_{k+1}+\u_{k}) 
    -\x_k||_2 \nonumber \\
    &= ||f_\theta(\z_{k+1}+\u_{k}+\boldsymbol{\eta}_k)
    - f_\theta(\z_{k+1}+\u_{k})||_2
    + \Delta_{\x_k}^{\boldsymbol{0}} \nonumber \\
    &\leq L||\boldsymbol{\eta}_k||_2
    + \Delta_{\x_k}^{\boldsymbol{0}},
\end{align}
where $L$ is the Lipschitz constant of $f_\theta$.

Similarly, for the $\u$-update we have
\begin{align}
    \Delta_{\u_k}^{\boldsymbol{\eta}}
    &= ||\u_{k+1}-\u_k||_2 \nonumber \\
    &= ||(\u_k+\z_{k+1}-\x_{k+1})-\u_k||_2 \nonumber \\
    &= ||\z_{k+1}-\x_{k+1}||_2 \nonumber \\
    &= ||\z_{k+1}-f_\theta(\z_{k+1}+\u_k +\boldsymbol{\eta}_k) ||_2 \nonumber \\
    &= ||\z_{k+1} - f_\theta(\z_{k+1}+\u_k)
    + f_\theta(\z_{k+1}+\u_k) \notag\\
    &\quad - f_\theta(\z_{k+1}+\u_k +\boldsymbol{\eta}_k) ||_2 \nonumber \\
    &\leq ||\z_k - f_\theta(\z_{k+1}+\u_k) ||_2 \notag\\
    &\quad + ||f_\theta(\z_{k+1}+\u_k) - f_\theta(\z_{k+1}+\u_k +\boldsymbol{\eta}_k) ||_2 \nonumber \\
    &= \Delta_{\u_k}^{\boldsymbol{0}}
    + ||f_\theta(\z_{k+1}+\u_k) - f_\theta(\z_{k+1}+\u_k +\boldsymbol{\eta}_k) ||_2 \nonumber \\
    &\leq \Delta_{\u_k}^{\boldsymbol{0}} + L||\boldsymbol{\eta}_k||_2
\end{align}
Hence, the additional divergence caused by noise injection at iteration $n$ is at most
\begin{equation}
\frac{2L}{\sqrt{P}}\|\boldsymbol{\eta}_k\|_2.
\end{equation}
Formally, the total deviation introduced across all iterations is bounded by
\begin{equation}
\sum_k \Delta_k^{\boldsymbol{\eta}} - \sum_k \Delta_k^{\boldsymbol{0}}
\;\leq\; \frac{2L}{\sqrt{P}} \sum_k \|\boldsymbol{\eta}_k\|_2.
\end{equation}
The convergence is preserved as long as the cumulative effect of noise injection remains finite, $\sum_{k=1}^\infty \|\boldsymbol{\eta}_k\|_2 < \infty$, and the noise injection is diminishing. This condition can be enforced by an appropriate noise scheduling strategy, e.g., by decaying the noise magnitude over iterations. Then the additional divergence introduced by noise injection remains bounded, ensuring that the overall convergence behavior of the algorithm is unaffected.
\end{proof}

\input{Tables/table_params}

\section{Experimental Details} \label{appdx:details}

\subsection{Implementation Details} \label{appdx:imp_details}
We implemented our method following \algoref{alg:proposed_admm}. For the JPEG restoration problem, we used the standard non-differentiable measurement, but employed a differentiable surrogate forward operator~\citep{reich2024differentiable} to enable gradient computation. For the inpainting task, we initialized our algorithm with an additional median filter, consistent with~\cite{garber2025CM4IR}. For phase retrieval, we obtained the measurements via $\y = |\mathcal{F}(\mathcal{M} \odot \x)|$, with $\mathcal{M}$ denoting the coded mask, $\mathcal{F}$ the Fourier transform, and $|\cdot|$ is entry-wise magnitude, similar to~\citep{xu2024_DPnP}. We note that phase retrieval suffers from ambiguity since magnitude-only measurements do not preserve phase information, resulting in multiple mathematically valid channel inversions. To identify the correct reconstruction, all eight inversions were evaluated, and the inversion minimizing LPIPS to a single pre-generated unconditional CM model output for the dataset was chosen. Note this does not increase the total NFE, since it is only performed at the output.

As described in \secref{sec:final_algo}, the data fidelity was implemented using an SVD-based approach for small-scale linear tasks (\eg, super-resolution, Gaussian deblurring, inpainting). For MRI reconstruction, we employed conjugate gradient (CG), which is well suited for the large-scale linear systems arising in physics-driven reconstruction~\citep{knoll2020deep-survey, yaman2020SSDU, yaman2021_3D-LGE, demirel2023SIIM, alcalar2024_ISBI, alcalar2025_SPIC_SSDU}. For nonlinear inverse problems, first-order optimization methods were used, with iterations run until a convergence tolerance was met.

\subsection{Hyperparameter Settings} \label{appdx:hyperparameters}

\paragraph{Noise schedule} We adopt the noise scheduling strategy of CM4IR~\cite{garber2025CM4IR}, which provides a compact parametrization requiring only a small number of hyperparameters while ensuring a stable, monotonic decrease in the injected noise across iterations. Following standard DDPM notation, the forward noising process uses a monotonically increasing variance schedule $\{\beta_i\}_{i=1}^{\mathrm{T}} \subset (0,1)$, with  $\alpha_i = 1 - \beta_i$, $\bar{\alpha}_i = \prod_{j=1}^{i}\alpha_{j}$. In CM4IR, two hyperparameters are used to define the noising process through these DDPM parameters. First, an initial diffusion index $i_\mathrm{N} \in [1,1000]$ selects the initial perturbation level as $t_\mathrm{N} = \sqrt{1 - \bar{\alpha}_{i_\mathrm{N}}}.$
Then a decay rate $\gamma$ is used to define the subsequent perturbation levels as $t_{\mathrm{N}-k} = \sqrt{1-(1+\gamma)^k\bar{\alpha}_{i_\mathrm{N}}}.$
Thus, a larger $\gamma$ corresponds to a faster decay in the injected noise across sampling iterations, while a larger $i_N$ increases the initial perturbation level from which the trajectory begins. We also follow the observation that CM noise level should slightly exceed the perturbation level to reflect the inherent measurement noise in the inverse problem, and use the same per-iteration offset vector $\{t_n\}_{n=1}^{\mathrm{N}}>0$ as in CM4IR with CM noise level as $(1+\delta_n) \, t_n$. 

\paragraph{Penalty parameter} For ADMM, the penalty parameter must stay positive and typically increase across iterations to progressively tighten the consistency constraint between the primal and dual variables. To reduce the dimensionality of this hyperparameter space, we parametrize the full sequence using a softplus-transformed linear interpolation:
$\rho_t = \mathrm{Softplus}\big(\mathrm{Linspace}(\rho_\mathrm{start}, \rho_\mathrm{end}, \mathrm{N})\big)$
This ensures positivity, provides smooth growth across iterations, and requires only two user-set endpoints $(\rho_\mathrm{start}, \rho_\mathrm{end})$.

\paragraph{Momentum coefficient} To improve stability and avoid oscillation, we use a simple decaying schedule for the momentum terms. The user specifies an initial coefficient $\mu_\mathrm{N}$, and subsequent coefficients are scaled down by the iteration index: $\mu_n = \mu_\mathrm{N} / (\mathrm{N}-n+1)$. With this definition, the momentum gradually decreases toward the final iterations, reducing the risk of oscillatory behavior and preventing late-stage instability. 

All hyperparameter values associated with our main experimental setup are listed in ~\tabref{tab:parameters}. 

\subsection{Comparison Methods} \label{appdx:comparison}

\paragraph{DPS} We followed the official implementation provided by \citet{chung2023dps}. Since the likelihood weight was originally tuned for ImageNet and FFHQ, we re-tuned it for LSUN Bedroom and FastMRI knee datasets, using $\eta=1.0$ for natural images and $\eta=0.85$ for medical imaging. For CelebA-HQ, we used the hyperparameters given for FFHQ.

\paragraph{$\Pi$GDM} We relied on the official implementation reported by \citet{song2023pgdm}. For both CelebA-HQ and LSUN Bedroom, we tuned the likelihood weights for each nonlinear task to achieve the best performance, and used 100 NFEs with $\eta=0.0$.

\paragraph{DiffPIR} For DiffPIR, we used the implementation provided by \citet{zhu2023DiffPIR}. For CelebA-HQ, we adopted the hyperparameters reported in Appendix B for the FFHQ dataset. Similarly, for LSUN Bedroom, we begin with the ImageNet hyperparameters from the same table and fine-tune them to achieve optimal performance on LSUN Bedroom.

\paragraph{DPnP} We followed the official implementation provided by~\citet{xu2024_DPnP}. For the CelebA-HQ dataset, we initially adopted the hyperparameters used for the FFHQ dataset, as described in Appendix D, and subsequently adjusted them as necessary to achieve optimal performance. For the LSUN Bedroom dataset, we began with the same hyperparameter settings used for CelebA-HQ and further refined them to obtain the best results.

\paragraph{PnP-DM} The official implementation as specified by \citet{wu2024_PnP-DM} was employed. Hyperparameters for CelebA-HQ were kept as originally reported. For LSUN Bedroom, the noise level schedule parameter $\beta_d$ was set to 14.9, which we found to achieve the best reconstruction performance. Although the codebase contains solvers for Gaussian deblurring and super-resolution, inpainting functionality was added through our own implementation, and its parameters were tuned to achieve optimal performance.

\paragraph{DDS} We used the official implementation of \citet{chung2024decomposed} with $\gamma=1.0$ and $\eta=0.85$. In contrast to clinical MRI reconstruction, where subsampled k-space measurements already contain inherent noise and no clean reference is available, the original method adds Gaussian noise to clean data. To better reflect the clinical setting, we applied DDS directly on the noisy subsampled k-space measurements. We tuned the CG steps depending on the data SNR.

\paragraph{CM} The official public repositories provided by \citet{song2023_consistency_models} were followed during the implementation. While the codebase includes iterative functions for inpainting and super-resolution, it does not provide one for Gaussian deblurring, which we implemented ourselves. We applied 40 steps for all inverse problems as advised by the authors.

\paragraph{CoSIGN} We used the official implementation provided by \citet{zhao2024cosign}. Since ControlNet training checkpoints are not available for inpainting (70\%) and Gaussian Deblurring tasks, we trained them from scratch. We used the provided checkpoint for super-resolution ($\times4$). 

\paragraph{CM4IR} For CM4IR, we relied on the official implementation and hyperparameter settings provided by \citet{garber2025CM4IR}. Since these configurations were extensively tuned by the authors, we retained them without modification. We note that the public implementation of CM4IR contains a heuristic Tikhonov-type stabilization term in its pseudoinverse calculation for highly ill-conditioned problems, which is not discussed in~\citep{garber2025CM4IR}. Nonetheless, we incorporate this term with the suggested values. 

\section{Training Details of Diffusion and Consistency Models} \label{apx:DM_CM_4_MRI}

We train the CMs on CelebA-HQ and fastMRI following the publicly available codes from~\citep{karras2022elucidating,song2023_consistency_models}. For both datasets, we first train an EDM-based DM~\citep{karras2022elucidating} and then distill it into a robust CM. Notably, for fastMRI, we establish, for the first time, a CM specifically designed for MRI data.  

We adopt the same U-Net hyperparameters as used for the LSUN Bedroom networks~\citep{song2023_consistency_models}, except for the number of input and output channels for fastMRI. For CelebA-HQ, the DM is trained for 3.55M iterations with a batch size of 16, and subsequently distilled into a CM using LPIPS loss for 1.175M iterations with a batch size of 32. For fastMRI, the DM is trained for 1M iterations with a batch size of 4, and then distilled into a CM using $\ell_2$ loss for 700k iterations with the same batch size.

\section{Further Experimental Results}

\input{Tables/momentum_vs_NFE_ablation}
\input{Tables/result_natural_NFE_sigma_y_celeba}

\subsection{Effect of Momentum Across NFEs} \label{appdx:second_ablation}
We study the effect of momentum across different numbers of NFEs. As shown in Table~\ref{tab:momentum_ablation}, the gains from momentum are modest at higher NFEs, which is consistent with standard ADMM behavior where convergence is already well-established. However, in the low-NFE regime considered in this work, momentum provides a noticeable improvement in reconstruction quality. This suggests that while momentum is not the primary driver of performance, it plays an important role in accelerating convergence when only a small number of iterations is used.

\subsection{Hyperparameter Sensitivity Analysis} \label{appdx:sensitivity}

To assess the robustness of our chosen hyperparameters, we conducted a perturbation study on the Gaussian deblurring task using the CelebA-HQ dataset. Focusing on the case $\mathrm{N} = 4$, each of the nine hyperparameters given in \tabref{tab:parameters} was independently perturbed by up to $\pm 25\%$, applying multiplicative factors drawn from a uniform distribution. For every image in the dataset, a new perturbed hyperparameter set was sampled and the full reconstruction process was rerun. We then compared the reconstruction quality against the baseline configuration.

\begin{figure}[!b]
    \centering
    \includegraphics[width=1.0\columnwidth]{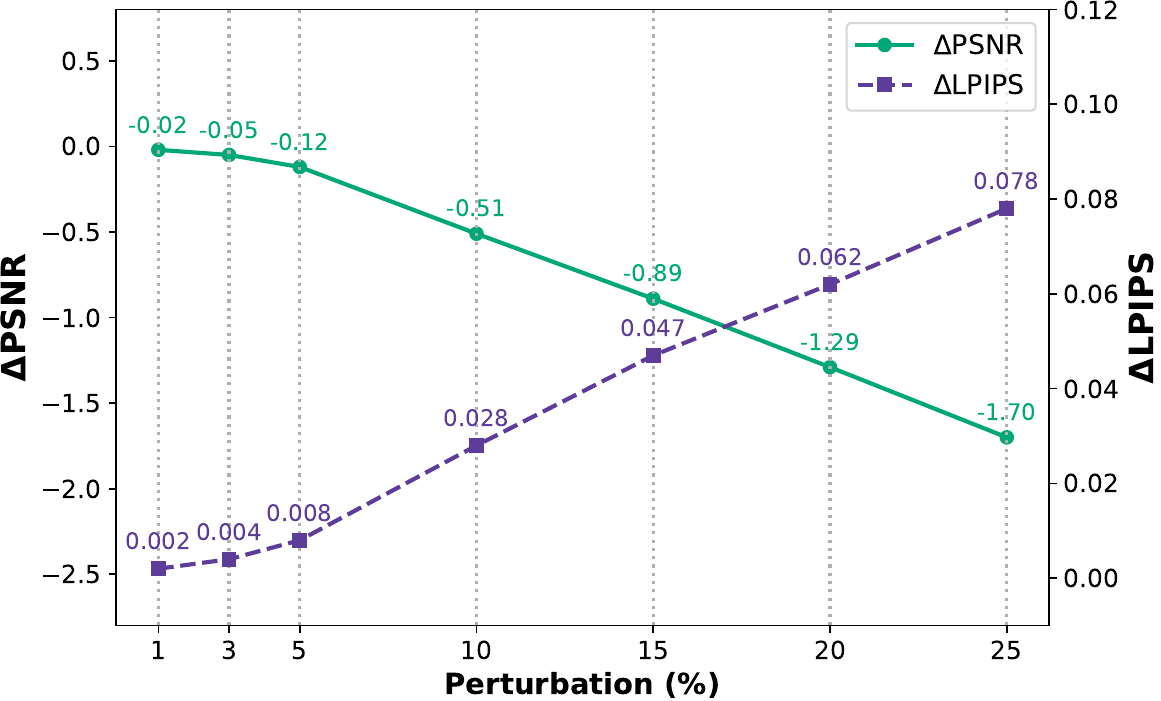} 
    \caption{Absolute change in PSNR and LPIPS when perturbing the reported PnP-CM hyperparameters. Even with up to 25\% variation, the deviations remain small, indicating strong robustness.}
    \vspace{-4mm}
    \label{fig:sensitivity}
\end{figure}

\figref{fig:sensitivity} summarizes the effect of increasing perturbation magnitude on performance by reporting the \emph{absolute} changes. As the perturbation level grows, both metrics degrade gradually, with PSNR decreasing and LPIPS increasing in a smooth and monotonic fashion. Importantly, the magnitude of these changes remains small: even at $25\%$ hyperparameter perturbation, the performance drop is modest and the algorithm continues to produce stable and visually plausible reconstructions.

\vspace{1mm}
\subsection{Performance Across Different Noise Levels and Sampling Steps} \label{appdx:diff_nfe_and_noise}
\tabref{tab:natural_nfe_sigmay} reports extended quantitative results for varying noise levels and sampling steps on CelebA-HQ. PnP-CM maintains consistently strong performance across different inverse tasks, showing robustness to both the number of function evaluations and measurement noise. Increasing the sampling steps generally improves quality as expected, while performance remains competitive even in the few-step regime. 

\vspace{2mm}
\subsection{Additional Qualitative Results for Natural Image Datasets} \label{appdx:celeba_appdx}
\begin{figure*}[p]
    \centering
    \includegraphics[width=\textwidth]{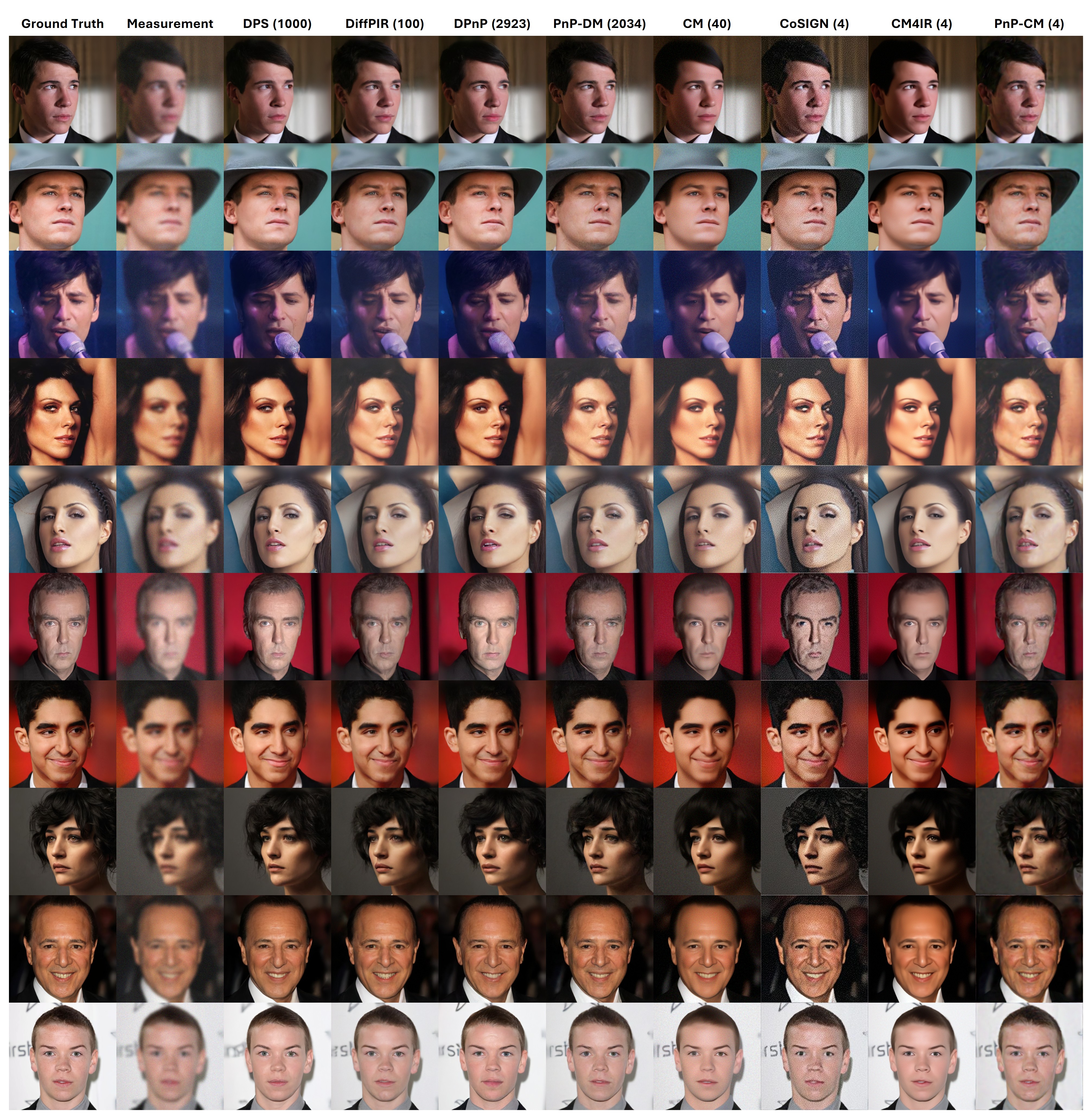} 
    \caption{Representative Gaussian deblurring results on CelebA-HQ with $\sigma_y = 0.05$. Unlike competing approaches, PnP-CM better preserves fine-scale facial structures and recovers sharper, more realistic textures.}
    \label{fig:clbA_gauss_apx}
\end{figure*}

\begin{figure*}[p]
    \centering
    \includegraphics[width=\textwidth]{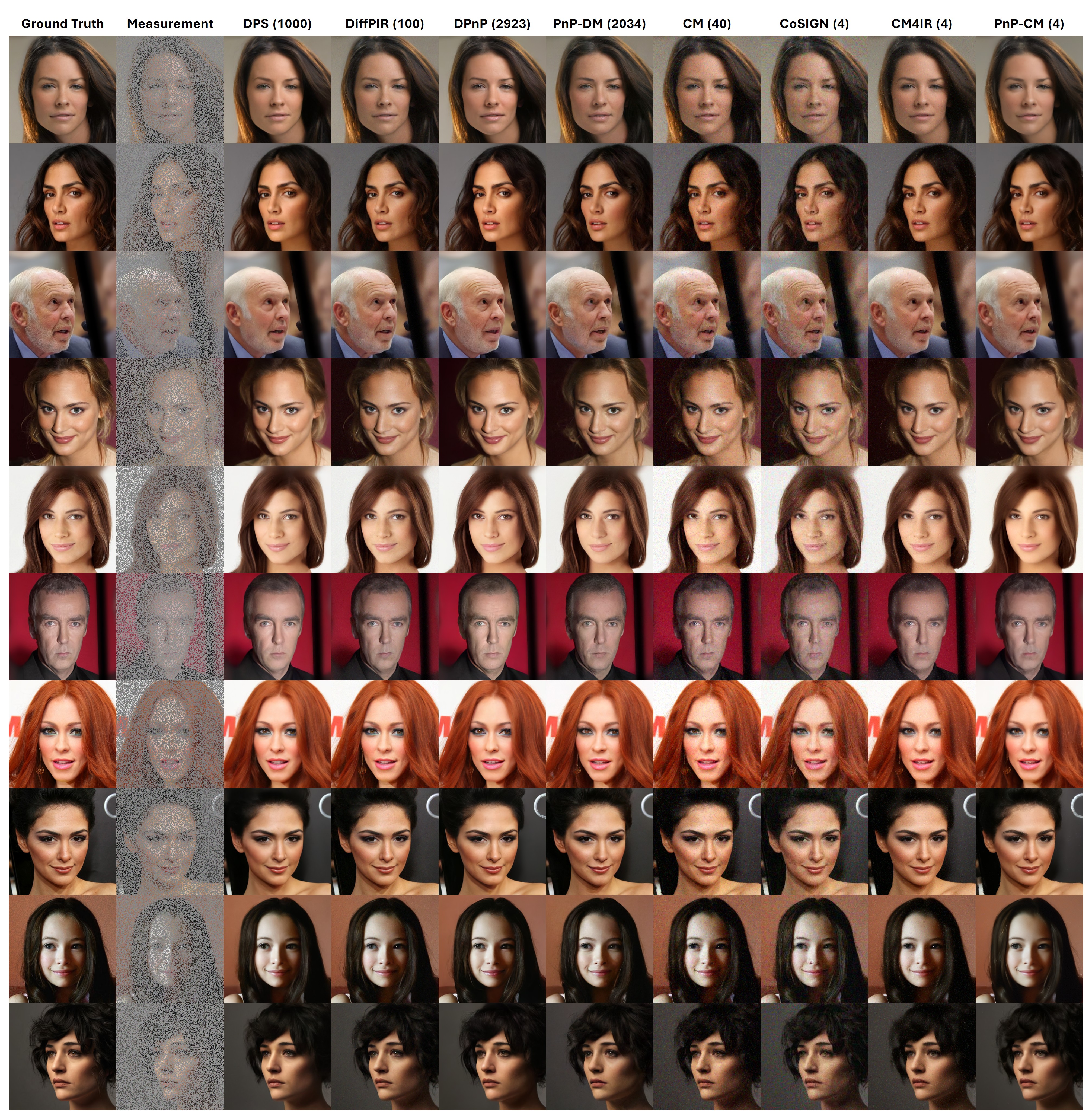}
    \caption{Demonstration of inpainting results on CelebA-HQ with $\sigma_y = 0.05$. PnP-CM reconstructs the missing regions with coherent facial geometry and realistic textures, producing completions that align more closely with the ground truth than competing methods.}
    \label{fig:clbA_inp_apx}
\end{figure*}

\begin{figure*}[p]
    \centering
    \includegraphics[width=\textwidth]{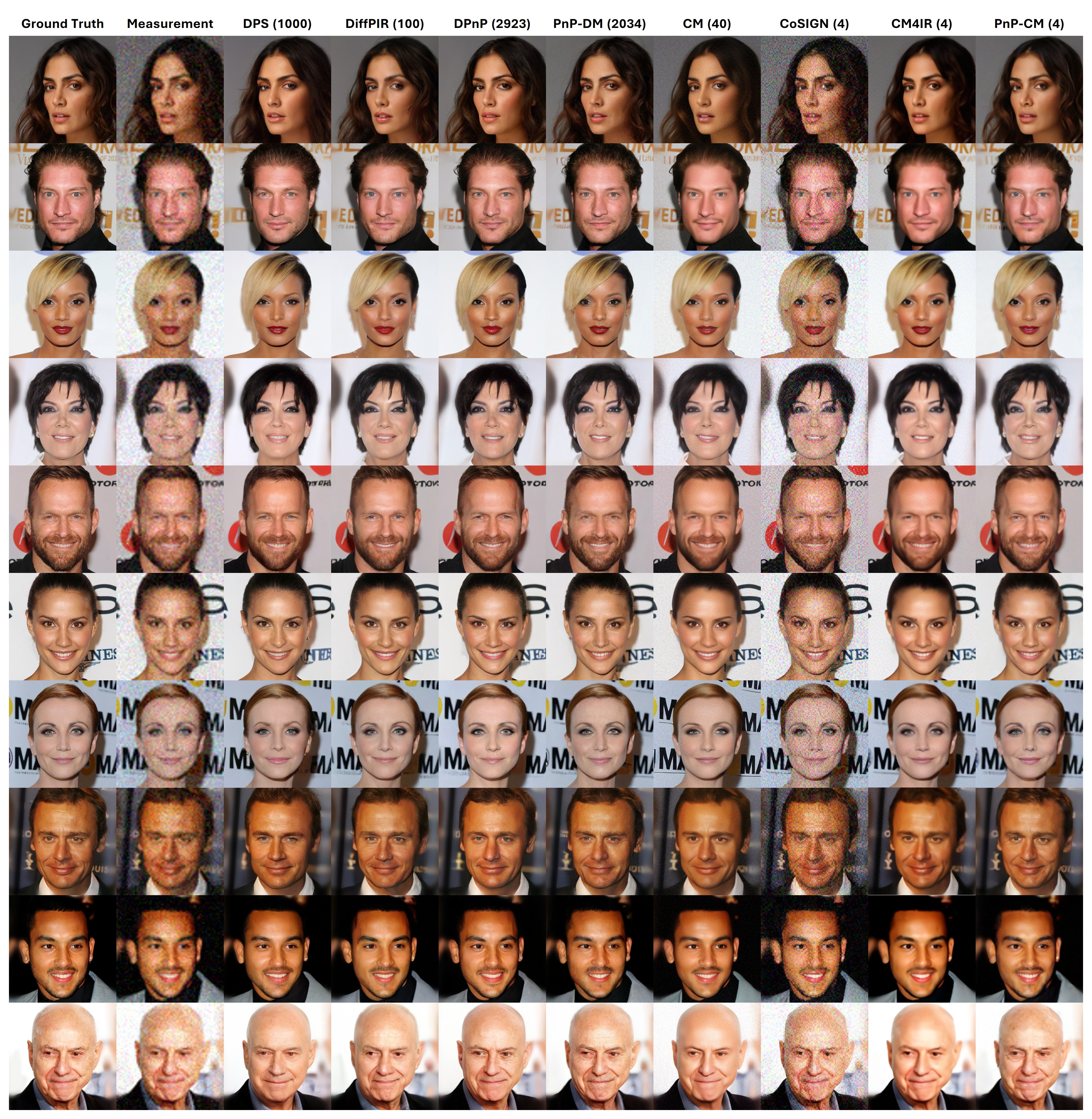}
    \caption{Illustrative $4\times$ super-resolution results on CelebA-HQ with $\sigma_y = 0.05$. PnP-CM restores high-frequency details with improved clarity and denoising, yielding visually richer reconstructions.}
    \label{fig:clbA_sr_apx}
\end{figure*}

Alongside the results presented in the main paper, we include further qualitative comparisons on CelebA-HQ for two representative inverse problems: Gaussian deblurring and $4\times$ super-resolution. As shown in Figs.~\ref{fig:clbA_gauss_apx}, \ref{fig:clbA_inp_apx}, and \ref{fig:clbA_sr_apx}, PnP-CM consistently reconstructs sharper facial details and more realistic high-frequency textures compared to all baseline methods.

\begin{figure*}[p]
    \centering
    \includegraphics[width=\textwidth]{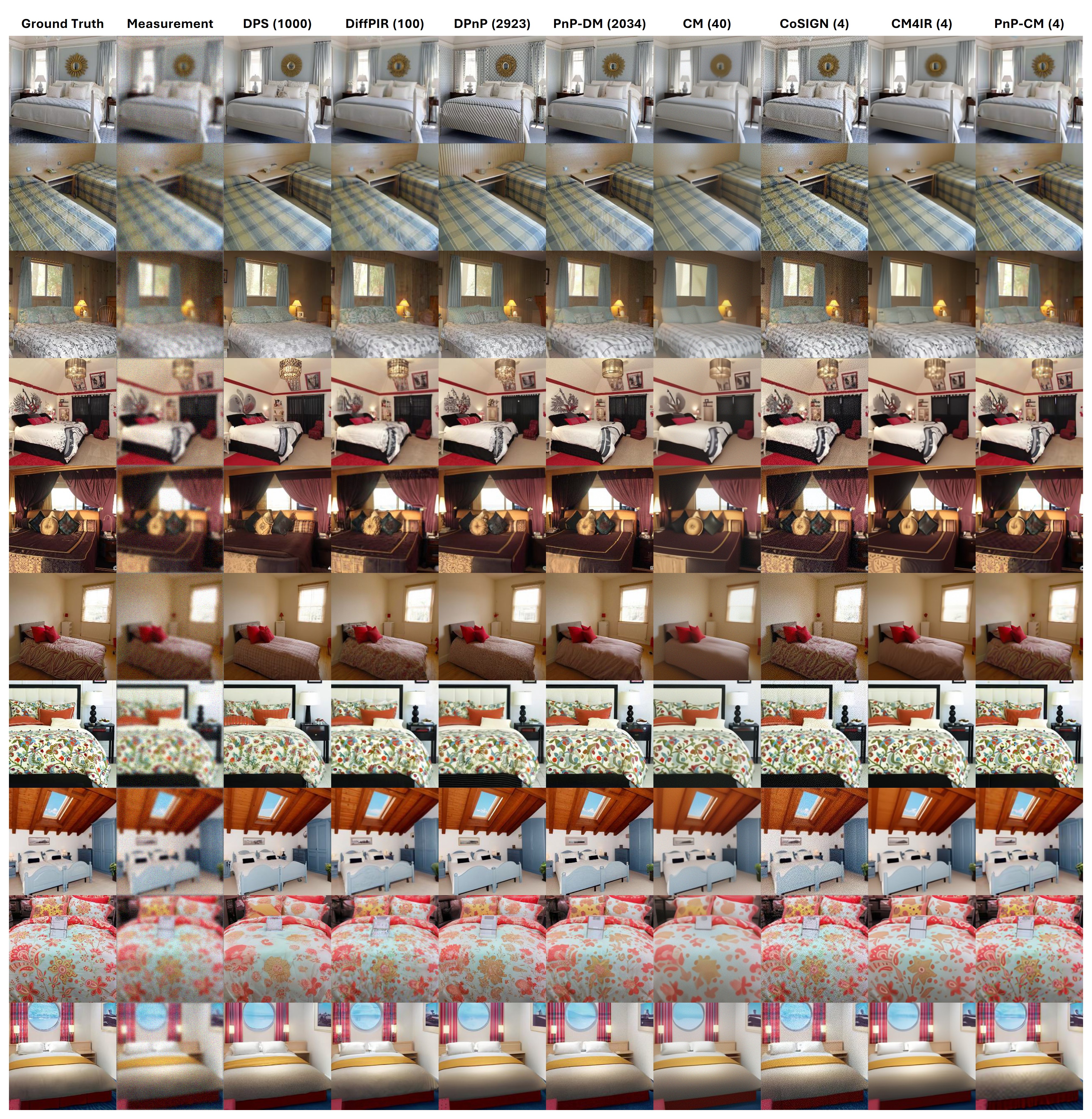} 
    \caption{Representative Gaussian deblurring results on LSUN Bedroom with $\sigma_y=0.05$. Comparisons with all baseline methods show that PnP-CM restores textures more faithfully and avoids oversmoothing.}
    \label{fig:gauss_apx}
\end{figure*}

\begin{figure*}[p]
    \centering
    \includegraphics[width=\textwidth]{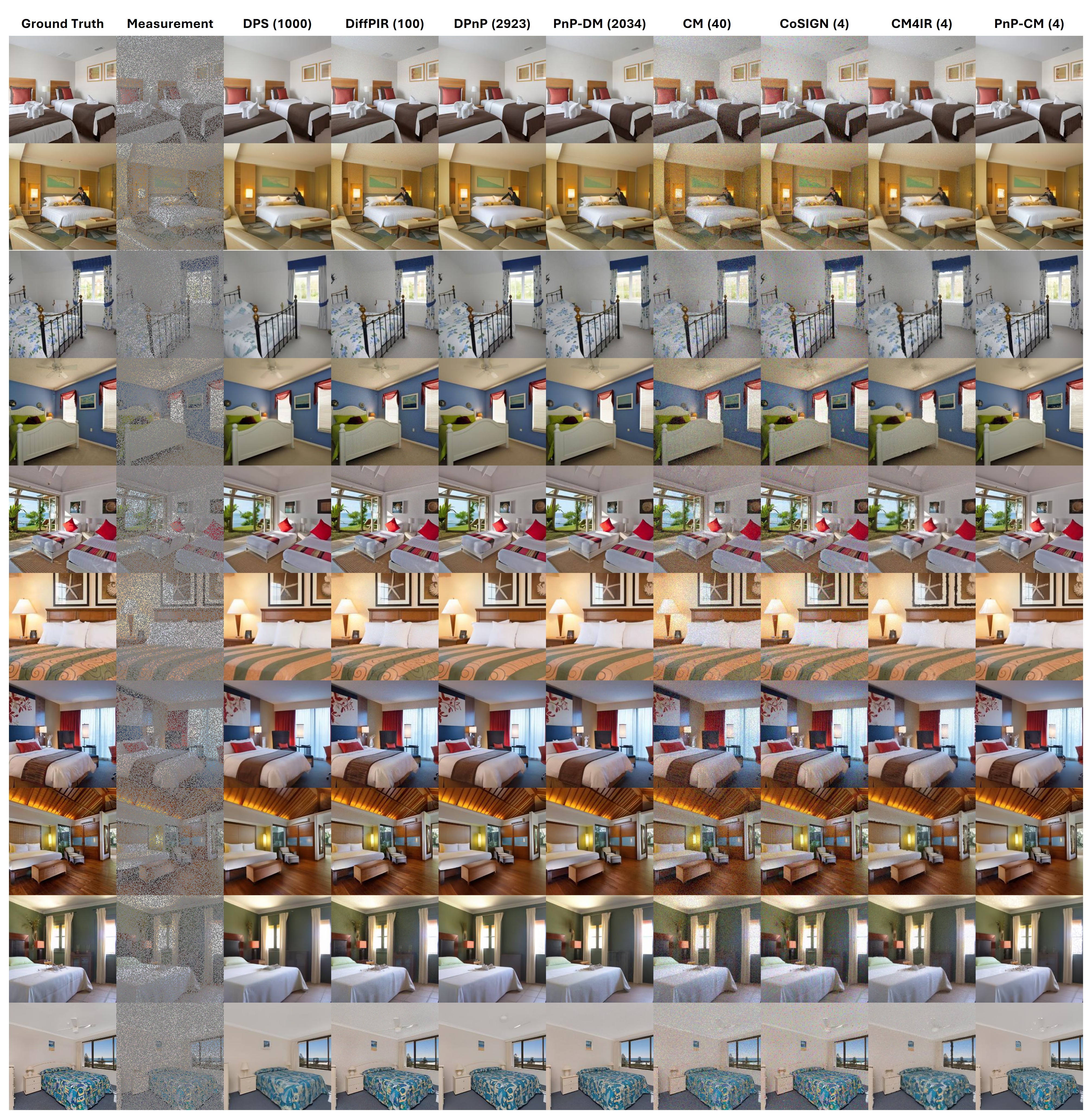} 
    \caption{Illustrative inpainting results on LSUN Bedroom with $\sigma_y=0.05$. Compared to other methods, PnP-CM fills missing regions with coherent structures and yields reconstructions closer to the ground truth.}
    \label{fig:inpainting_apx}
\end{figure*}

\begin{figure*}[p]
    \centering
    \includegraphics[width=\textwidth]{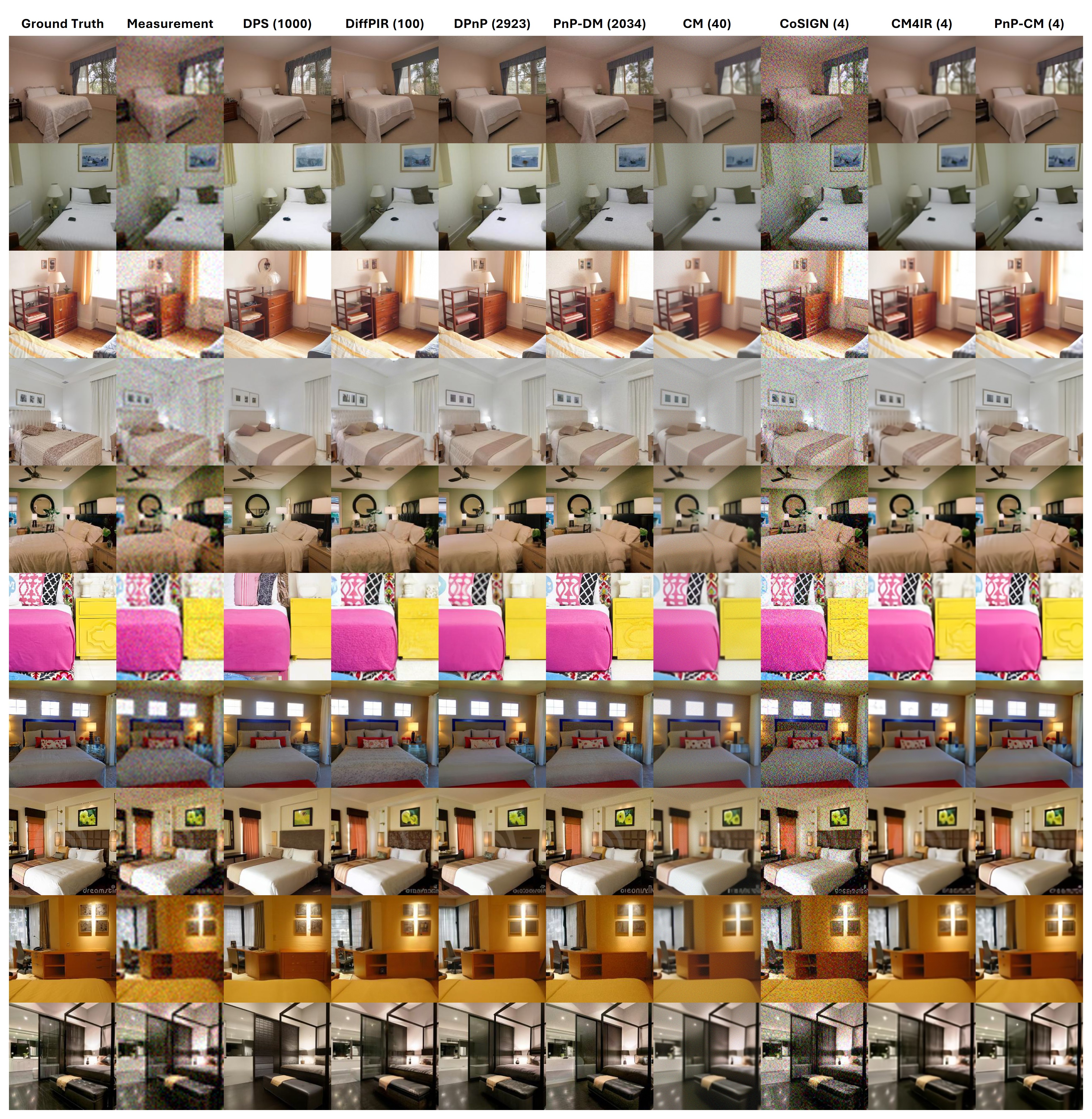} 
    \caption{Demonstration of super-resolution ($\times 4$) results on LSUN Bedroom with $\sigma_y=0.05$. Reconstructions are compared against all baseline methods, with PnP-CM producing sharper details and closer resemblance to the ground truth.}
    \label{fig:sr_apx}
\end{figure*}

For LSUN Bedroom, we also provide further qualitative results against all baseline methods discussed in the main text in Figs.~\ref{fig:gauss_apx}, \ref{fig:inpainting_apx}, and \ref{fig:sr_apx}, illustrating that the visual trends observed there hold across a wider set of examples.

\vspace{2mm}
\subsection{Additional Qualitative Results on MRI Reconstruction} \label{apx:MRI_recon}
\begin{figure*}[p]
    \centering
    \includegraphics[width=0.94\textwidth]{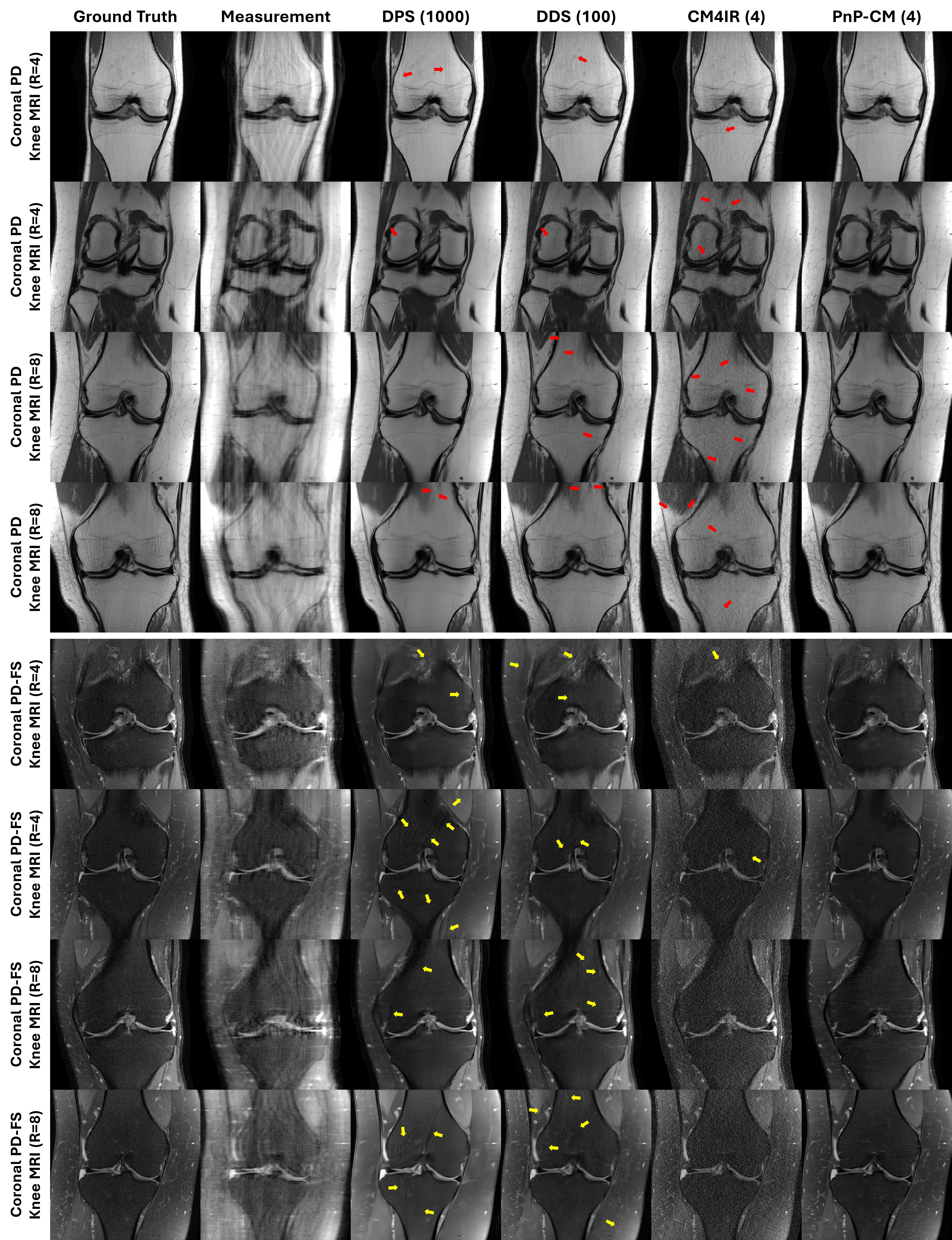} 
    \caption{Qualitative comparisons for $R = 4$ and $R = 8$ on the Coronal PD and Coronal PD-FS datasets across different methods. The proposed method, PnP-CM, consistently demonstrates superior performance by effectively reducing artifacts.}
    \label{fig:appdx_MRI}
\end{figure*}

We provide additional qualitative reconstruction results for coronal PD and PD-FS datasets at acceleration factors $R=4$ and $R=8$, with representative examples from each case shown in \figref{fig:appdx_MRI}.

%% file: Tables/table_params.tex
\begin{table*}[t]
\caption{Complete set of hyperparameters used across all tasks in this study, including the noise schedule, ADMM penalty parameters, and momentum coefficients, for both natural image experiments and medical imaging applications.}
\vspace{-4mm}
\label{tab:parameters}
\begin{center}
\begin{small}
\renewcommand{\arraystretch}{1.05}
\setlength{\tabcolsep}{7.9pt}

\begin{tabular}{lccccc}
\toprule

% ============================================================
% CELEBA-HQ
% ============================================================
\rowcolor{gray!20}
\multicolumn{6}{c}{\textbf{Inverse Problems on Natural Images (CelebA-HQ~\cite{karras2018progressive})}} \\
\midrule

\addlinespace[2pt]
SR $\times$4 & $i_\textrm{N}=150$ & $\gamma=0.2$ & $\delta_n=(0.3,0.05,0.2,0.2)$ & $\rho=(-4.0,4.0)$ & $\mu_\textrm{N}=0.2$  \\
\arrayrulecolor{gray!40}
\addlinespace[2pt]\midrule
\addlinespace[3pt]

\addlinespace[2pt]
Gaussian Deblur & $i_\textrm{N}=100$ & $\gamma=0.1$ & $\delta_n=(0.3,0.2,0.1,0.1)$ & $\rho=(-4.0,6.5)$ & $\mu_\textrm{N}=0.1$  \\
\arrayrulecolor{gray!40}
\addlinespace[2pt]\midrule
\addlinespace[3pt]

\addlinespace[2pt]
Inpainting & $i_\textrm{N}=100$ & $\gamma=0.1$ & $\delta_n=(0.3,0.2,0.8,0.8)$ & $\rho=(-6.0,12.0)$ & $\mu_\textrm{N}=0.05$  \\
\arrayrulecolor{gray!40}
\addlinespace[2pt]\midrule
\addlinespace[3pt]

\addlinespace[2pt]
JPEG Restoration & $i_\textrm{N}=250$ & $\gamma=0.3$ & $\delta_n=(0.1,0.1,0.2,4.0)$ & $\rho=(-2.5,2.5)$ & $\mu_\textrm{N}=0.05$  \\
\arrayrulecolor{gray!40}
\addlinespace[2pt]\midrule
\addlinespace[3pt]

\addlinespace[2pt]
Nonlinear Deblur & $i_\textrm{N}=250$ & $\gamma=0.15$ & $\delta_n=(0.05,0.05,0.05,0.1)$ & $\rho=(-2.0,0.0)$ & $\mu_\textrm{N}=0.05$  \\
\arrayrulecolor{gray!40}
\addlinespace[2pt]\midrule
\addlinespace[3pt]

\addlinespace[2pt]
Phase Retrieval & $i_\textrm{N}=250$ & $\gamma=0.03$ & {\footnotesize $\delta_n=(0.1,0.1,0.1,0.2,0.2,0.2,0.2,0.2)$} & $\rho=(-2.0,0.0)$ & $\mu_\textrm{N}=0.05$  \\
\arrayrulecolor{black}\midrule

% ============================================================
% MRI SECTION HEADER
% ============================================================
\rowcolor{gray!20}
\multicolumn{6}{c}{\textbf{Medical Imaging (MRI Reconstruction using fastMRI~\cite{knoll2020fastmri_dataset-journal})}} \\
\midrule

\addlinespace[2pt]
Coranal PD -- (R=4) & $i_\textrm{N}=50$ & $\gamma=0.1$ & $\delta_n=(0.3,2.0,6.0,2.5)$ & $\rho=(-4.5,-1.0)$ & $\mu_\textrm{N}=0.20$  \\
\arrayrulecolor{gray!40}
\addlinespace[2pt]\midrule
\addlinespace[3pt]

\addlinespace[2pt]
Coranal PD -- (R=8) & $i_\textrm{N}=50$ & $\gamma=0.1$ & $\delta_n=(0.35,3.5,7.5,3.5)$ & $\rho=(-4.5,-1.5)$ & $\mu_\textrm{N}=0.50$  \\
\arrayrulecolor{gray!40}
\addlinespace[2pt]\midrule
\addlinespace[3pt]

\addlinespace[2pt]
Coranal PD-FS -- (R=4) & $i_\textrm{N}=50$ & $\gamma=0.1$ & $\delta_n=(0.2,3.0,4.0,2.5)$ & $\rho=(-2.5,-0.5)$ & $\mu_\textrm{N}=0.05$  \\
\arrayrulecolor{gray!40}
\addlinespace[2pt]\midrule
\addlinespace[3pt]

\addlinespace[2pt]
Coranal PD-FS -- (R=8) & $i_\textrm{N}=50$ & $\gamma=0.1$ & $\delta_n=(0.4,9.5,4.5,1.5)$ & $\rho=(-2.5,0.5)$ & $\mu_\textrm{N}=0.45$  \\
\arrayrulecolor{black}\midrule

\end{tabular}
\end{small}
\end{center}
\vspace{-3ex}
\end{table*}

%% file: Tables/momentum_vs_NFE_ablation.tex
\begin{table}[!t]%[15]{r}{0.65\textwidth}
\captionsetup{type=table}
\caption{Effect of momentum across different NFEs on PD MRI (R=8). Momentum yields modest gains at higher NFEs, but provides noticeable improvements in the low-NFE regime.}
\vspace{-4mm}
\label{tab:momentum_ablation}
\begin{center}
\begin{footnotesize}
\renewcommand{\arraystretch}{0.8}
\setlength{\tabcolsep}{4.5pt}
\begin{tabular}{ccccc}
\toprule
\multicolumn{1}{c}{\textbf{Noise Inj.}} &
\multicolumn{1}{c}{\textbf{Moment.}} &
\multicolumn{1}{c}{\textbf{NFE=4}} &
\multicolumn{1}{c}{\textbf{NFE=10}} &
\multicolumn{1}{c}{\textbf{NFE=20}}\\
\midrule

\cmark & \xmark
    & 30.70 / 0.845 
    & 32.81 / 0.868 
    & 32.86 / 0.873 \\
\arrayrulecolor{gray!40}\midrule

\cmark & \cmark
    & 32.81 / 0.870 
    & 32.87 / 0.872 
    & 32.81 / 0.879 \\
\arrayrulecolor{black}\bottomrule
\end{tabular}
\end{footnotesize}
\end{center}
\vspace{-3mm}
\end{table}

%% file: Tables/result_natural_NFE_sigma_y_celeba.tex
\begin{table*}[ht]
\caption{Quantitative results on CelebA-HQ for different measurement noise levels ($\sigma_y$) and varying numbers of NFEs across super resolution ($\times 4$), Gaussian deblurring, and inpainting tasks.}
\vspace{-3ex}
\label{tab:natural_nfe_sigmay}
\begin{center}
\begin{small}
\renewcommand{\arraystretch}{1}
\setlength{\tabcolsep}{20pt}
\begin{tabular}{lccccc}
\toprule
\multicolumn{1}{l}{\textbf{NFE$\downarrow$}} & $\sigma_y$ &
\multicolumn{1}{c}{\textbf{SR $\times$4}} &
\multicolumn{1}{c}{\textbf{\shortstack{Gaussian \\ Deblurring}}} &
\multicolumn{1}{c}{\textbf{Inpainting}} \\ 
\cmidrule(lr){3-5}
 &  & {\scriptsize PSNR$\uparrow$} / {\scriptsize LPIPS$\downarrow$} &
      {\scriptsize PSNR$\uparrow$} / {\scriptsize LPIPS$\downarrow$} &
      {\scriptsize PSNR$\uparrow$} / {\scriptsize LPIPS$\downarrow$} \\
\midrule

\multirow{2}{*}{2} 
  & 0.025 
  & 27.44 / 0.272
  & 29.22 / 0.251
  & 28.77 / 0.213 \\
  
  & 0.05
  & 26.99 / 0.281
  & 28.78 / 0.256
  & 28.91 / 0.209 \\
\arrayrulecolor{gray!40}\midrule

\multirow{2}{*}{4} 
  & 0.025 
  & 27.78 / 0.265
  & 29.57 / 0.238
  & 29.14 / 0.206 \\
  
  & 0.05
  & 27.27 / 0.285
  & 28.94 / 0.249
  & 29.23 / 0.201 \\
\arrayrulecolor{gray!40}\midrule

\multirow{2}{*}{8} 
  & 0.025 
  & 27.91 / 0.264
  & 29.71 / 0.237
  & 29.23 / 0.205 \\
  
  & 0.05
  & 27.37 / 0.286
  & 29.05 / 0.250
  & 29.32 / 0.199 \\

\arrayrulecolor{black}\bottomrule

\end{tabular}
\end{small}
\end{center}
\vspace{-4ex}
\end{table*}